\documentclass[12pt,thmsa]{article}%
\usepackage{amsmath}
\usepackage{amssymb}
\usepackage{sw20jart}%
\usepackage{amsfonts}%
\usepackage{graphicx}
\providecommand{\U}[1]{\protect\rule{.1in}{.1in}}
\begin{document}

\title{AMBIGUOUS VOLATILITY AND ASSET PRICING IN CONTINUOUS TIME\thanks{Department of
Economics, Boston University, lepstein@bu.edu and School of Mathematics,
Shandong University, jsl@sdu.edu.cn. We gratefully acknowledge the financial
support of the National Science Foundation (awards SES-0917740 and 1216339),
the National Basic Research Program of China (Program 973, award 2007CB814901)
and the National Natural Science Foundation of China (award 10871118). We have
benefited also from very helpful comments by Pietro Veronesi and two referees,
and from discussions with Shige Peng, Zengjing Chen, Mingshang Hu, Gur
Huberman, Julien Hugonnier, Hiro Kaido, Jin Ma, Semyon Malamud, Guihai Zhao
and especially Jianfeng Zhang. Epstein is grateful also for the generous
hospitality of EIEF and CIRANO where much of this work was completed during
extended visits. The contents of this paper appeared originally under the
title \textquotedblleft Ambiguous volatility, possibility and utility in
continuous time,\textquotedblright\ first posted March 5, 2011. That paper now
focuses on a mathematically rigorous treatment of the model of utility and
excludes the application to asset pricing.}}
\author{Larry G. Epstein
\and Shaolin Ji}
\maketitle
\date{}

\begin{abstract}
This paper formulates a model of utility for a continuous time framework that
captures the decision-maker's concern with ambiguity about both volatility and
drift. Corresponding extensions of some basic results in asset pricing theory
are presented. First, we derive arbitrage-free pricing rules based on hedging
arguments. Ambiguous volatility implies market incompleteness that rules out
perfect hedging. Consequently, hedging arguments determine prices only up to
intervals. However, sharper predictions can be obtained by assuming preference
maximization and equilibrium. Thus we apply the model of utility to a
representative agent endowment economy to study equilibrium asset returns. A
version of the C-CAPM is derived and the effects of ambiguous volatility are described.

\medskip


\end{abstract}

\newpage

\section{Introduction}

\subsection{Objectives}

This paper formulates a model of utility for a continuous time framework that
captures the decision-maker's concern with ambiguity or model uncertainty. Its
novelty lies in the range of model uncertainty that is accommodated,
specifically in the modeling of ambiguity about both drift and volatility, and
in corresponding extensions of some basic results in asset pricing theory.
First, we derive arbitrage-free pricing rules based on hedging arguments.
Ambiguous volatility implies market incompleteness and thus, in general, rules
out perfect hedging. Consequently, hedging arguments determine prices only up
to intervals. However, sharper predictions can be obtained by assuming
preference maximization and equilibrium. Thus we apply the model of utility to
a representative agent endowment economy to study equilibrium asset returns in
a sequential Radner style market setup. A version of the C-CAPM is derived and
the effects of ambiguous volatility are described. A pivotal role for `state
prices' is demonstrated in both the hedging and equilibrium analyses thus
extending to the case of comprehensive ambiguity this cornerstone element of
asset pricing theory.

The model of utility is a continuous time version of multiple priors (or
maxmin) utility formulated by Gilboa and Schmeidler \cite{gs} for a static
setting. Related continuous time models are provided by Chen and Epstein
\cite{CE} and also Hansen, Sargent and coauthors (see Anderson et al.
\cite{ahs}, for example).\footnote{The discrete time counterpart of the former
is axiomatized in Epstein and Schneider \cite{es2003}.} In \emph{all} existing
literature on continuous time utility, ambiguity is modeled so as to retain
the property that all priors are equivalent, that is, they agree which events
are null. This universal restriction is driven not by an economic rationale
but rather by the technical demands of continuous time modeling, specifically
by the need to work within a probability space framework. Notably, in order to
describe ambiguity authors invariably rely on Girsanov's theorem for changing
measures. It provides a tractable characterization of alternative hypotheses
about the true probability law, but it also limits alternative hypotheses to
correspond to measures that are both mutually equivalent and that differ from
one another only in what they imply about drift. This paper defines a more
general framework within which one can model the utility of an individual who
is not completely confident in any single probability law for either drift or
volatility. This is done while maintaining a separation between risk aversion
and intertemporal substitution as in Duffie and Epstein \cite{DE}.

At a technical level, the analysis requires a significant departure from
existing continuous time modeling because ambiguous volatility cannot be
modeled within a probability space framework, where there exists a probability
measure that defines the set of null (or impossible) events. In our companion
paper Epstein and Ji \cite{eji}, we exploit and extend recent advances in
stochastic calculus that do not require a probability space framework. The
reader is referred to that paper for a rigorous treatment of the technical
details involved in defining a utility function that accommodates aversion to
ambiguity about volatility, including for proofs regarding utility, and also
for extensive references to the noted mathematics literature. Our treatment
below is less formal but is otherwise largely self-contained. Proofs are
provided here for all the asset pricing results.

\subsection{Why ambiguous volatility?\label{section-why}}

A large literature has argued that stochastic time varying volatility is
important for understanding empirical features of asset markets; for recent
examples, see Eraker and Shaliastovich \cite{eraker}, Drechsler
\cite{drechsler}, Bollerslev et al. \cite{bollerslev}, Bansal et al.
\cite{BKY}, Beeler and Campbell \cite{beeler}, Bansal et al. \cite{BKSY}, and
Campbell et al. \cite{campbell}, where the first three employ continuous time
models.\footnote{Bollerslev et al. argue extensively for the modeling
advantages of the continuous time framework. For example, they write that a
continuous time formulation \textquotedblleft has the distinct advantage of
allowing for the calculation of internally consistent model implications
across all sampling frequencies and return horizons.\textquotedblright} In
macroeconomic contexts, Bloom \cite{bloom} and Fernandez-Villaverde and
Guerron-Quintana \cite{fernandez} are recent studies that find evidence of
stochastic time varying volatility and its effects on real variables. In all
of these papers, evidence suggests that relevant volatilities follow
complicated dynamics. The common modeling response is to postulate
correspondingly complicated parametric laws of motion, including specification
of the dynamics of the volatility of volatility. However, one might question
whether agents in these models can learn these laws of motion precisely, and
more generally, whether it is plausible to assume that agents become
completely confident in any particular law of motion. In their review of the
literature on volatility derivatives, Carr and Lee \cite[pp. 324-5]{carrlee}
raise this criticism of assuming a particular parametric process for the
volatility of the underlying asset. The drawback they note is
\textquotedblleft the dependence of model value on the particular process used
to model the short-term volatility.\textquotedblright\ They write that
\textquotedblleft the problem is particularly acute for volatility models
because the quantity being modeled is not directly observable. Although an
estimate for the initially unobserved state variable can be inferred from
market prices of derivative securities, noise in the data generates noise in
the estimate, \textit{raising doubts that a modeler can correctly select any
parametric stochastic process from the menu of consistent alternatives}%
.\textquotedblright

Thus we are led to develop a model of preference that accommodates ambiguity
about volatility. In the model the individual takes a stand only on bounds
rather than on any particular parametric model of volatility dynamics. Thus
maximization of preference leads to decisions that are robust to
misspecifications of the dynamics of volatility (as well as drift).
Accordingly, we think of this aspect of our model as providing a way to
\textit{robustify stochastic volatility modeling}.

To illustrate the latter perspective, consider a stochastic environment with a
one-dimensional driving process. By a stochastic volatility model we mean the
hypothesis that the driving process has zero drift and that its volatility is
stochastic and is described by a single process $\left(  \sigma_{t}\right)  $.
The specification of a single process for volatility indicates the investor's
complete confidence in the implied dynamics. Suppose, however, that $\left(
\sigma_{t}^{1}\right)  $ and $\left(  \sigma_{t}^{2}\right)  $ describe two
alternative stochastic volatility models that are put forth by expert
econometricians; for instance, they might conform to the Hull and White
\cite{hw} and Heston \cite{heston} parametric forms respectively. The models
have comparable empirical credentials and are not easily distinguished
empirically, but their implications for optimal choice (or for the pricing of
derivative securities, which is the context for the earlier quote from Carr
and Lee) differ significantly. Faced with these two models, the investor might
place probability $\frac{1}{2}$ on each being the true model. But why should
she be certain that either one is true? Both $\left(  \sigma_{t}^{1}\right)  $
and $\left(  \sigma_{t}^{2}\right)  $ may fit data well to some approximation,
but other approximating models may do as well. An intermediate model such as
$\left(  \frac{1}{2}\sigma_{t}^{1}+\frac{1}{2}\sigma_{t}^{2}\right)  $ is one
alternative, but there are many others that \textquotedblleft lie
between\textquotedblright\ $\left(  \sigma_{t}^{1}\right)  $ and $\left(
\sigma_{t}^{2}\right)  $ and that plausibly should be taken into account.
Accordingly, we are led to hypothesize that the investor views as possible all
volatility processes with values lying in the interval $[\underline{\sigma
}_{t}\left(  \omega\right)  ,\overline{\sigma}_{t}\left(  \omega\right)  ]$
\ for every $t$ and $\omega$, where%
\begin{equation}
\underline{\sigma}_{t}\left(  \omega\right)  =\min\{\sigma_{t}^{1}\left(
\omega\right)  ,\sigma_{t}^{2}\left(  \omega\right)  \}\text{ and }%
\overline{\sigma}_{t}\left(  \omega\right)  =\max\{\sigma_{t}^{1}\left(
\omega\right)  ,\sigma_{t}^{2}\left(  \omega\right)  \}\text{.}%
\label{robustvol}%
\end{equation}
Given also the conservative nature of multiple priors utility, the individual
will be led thereby to take decisions that are robust to (many)
misspecifications of the dynamics of volatility. This special case of our
model is described further in Section \ref{section-priors}.

A possible objection to modeling ambiguity about volatility might take the
form: \textquotedblleft One can approximate the realized quadratic variation
of a stock price (for example) arbitrarily well from frequent observations
over any short time interval, and thus estimate the law of motion for its
volatility extremely well. Consequently, ambiguity about volatility is
implausible for a sophisticated agent.\textquotedblright\ However, even if one
accepts the hypothesis that, contrary to the view of Carr and Lee, accurate
estimation is possible, such an objection relies also on the assumption of a
tight connection between the past and future that we relax. We are interested
in situations where realized past volatility may not be a reliable predictor
of volatility in the future. The rationale is that the stochastic environment
is often too complex for a sophisticated individual to believe that her
theory, whether of volatility or of other variables, captures all aspects.
Being sophisticated, she is aware of the incompleteness of her theory.
Accordingly, when planning ahead she believes there may be time-varying
factors excluded by her theory that she understands poorly and that are
difficult to identify statistically. Thus she perceives ambiguity when looking
into the future. The amount of ambiguity may depend on past observations, and
may be small for some histories, but it cannot be excluded \textit{a priori}.

A similar rationale for ambiguity is emphasized by Epstein and Schneider
\cite{es2008,esAR}. Nonstationarity is emphasized by Ilut and Schneider
\cite{ilut} in their model of business cycles driven by ambiguity. In finance,
Lo and Mueller \cite{lo} argue that the (perceived) failures of the dominant
paradigm, for example, in the context of the recent crisis, are due to
inadequate attention paid to the kind of uncertainty faced by agents and
modelers. Accordingly, they suggest a new taxonomy of uncertainty that extends
the dichotomy between risk and ambiguity (or `Knightian uncertainty'). In
particular, they refer to \textit{partially reducible uncertainty} to describe
\textquotedblleft situations in which there is a limit to what we can deduce
about the underlying phenomena generating the data. Examples include
data-generating processes that exhibit: (1) stochastic or time-varying
parameters that vary too frequently to be estimated accurately; (2)
nonlinearities too complex to be captured by existing models, techniques and
datasets; (3) nonstationarities and non-ergodicities that render useless the
Law of Large Numbers, Central Limit Theorem, and other methods of statistical
inference and approximation; and (4) the dependence on relevant but unknown
and unknowable conditioning information.\textquotedblright\ Lo and Mueller do
not offer a model. One can view this paper as an attempt to introduce some of
their concerns into continuous time modeling and particularly into formal
asset pricing theory.

The natural question is whether and in what form the cornerstones of received
asset pricing theory extend to a framework with ambiguous volatility. Some
initial steps in answering this question are provided in Section
\ref{section-asset}.\footnote{Early work on the pricing of derivative
securities when volatility is ambiguous includes Lyons \cite{Lyons} and
Avellaneda et al. \cite{ALP}. See Section \ref{section-asset} for the relation
to our analysis and for additional references.} A notable finding is that both
equilibrium and \textquotedblleft no-arbitrage' asset prices can be
characterized by means of `state prices' even though the analysis cannot be
undertaken in a probability space framework (which precludes talking about
state price \emph{densities} or about \emph{equivalent} martingale, or risk
neutral, measures). First, however, the remainder of the introduction provides
an informal outline of our approach to modeling ambiguous volatility. Then
Section \ref{section-utility} describes the new model of utility. Following
the asset pricing results, concluding remarks are offered in Section
\ref{section-conclude}. Proofs are collected in appendices.

\subsection{An informal outline\label{section-trinomial}}

Time varies over $\{0,h,2h,...,\left(  n-1\right)  h,nh\}$, where $0<h<1$
scales the period length and $n$ is a positive integer with $nh=T$.
\ Uncertainty is driven by the colors of balls drawn from a sequence of urns.
It is known that each urn contains 100 balls that are either red ($R$), green
($G$), or yellow ($Y$), and that the urns are constructed independently
(informally speaking). A ball is drawn from each urn and the colors drawn
determine the evolution of the state variable $B=\left(  B_{t}\right)  $
according to: $B_{0}=0$ and, for $t=h,...,nh$,
\[
dB_{t}\equiv B_{t}-B_{t-h}=\left\{
\begin{array}
[c]{cc}%
h^{1/2} & \text{if }R_{t}\\
-h^{1/2} & \text{if }G_{t}\\
0 & \text{if }Y_{t}%
\end{array}
\right.
\]
We describe three alternative assumptions regarding the additional information
available about the urns. They provide intuition for continuous time models
where (respectively) the driving process is (i) a standard Brownian motion,
(ii) a Brownian motion modified by ambiguous drift, and (iii) a Brownian
motion modified by ambiguous volatility. The first two are included in order
to provide perspective on the third.

\medskip

\noindent Scenario 1: You are told further that $Y=0$ and that $R=G$ for each
urn (thus all urns are known to have the identical composition). The state
process $\left(  B_{t}\right)  $ can be described equivalently in terms of the
measure $p_{0}=\left(  \frac{1}{2},\frac{1}{2},0\right)  $ and its i.i.d.
product that induces a measure $P_{0}$ on trajectories of $B$. Thus we have a
random walk that, by Donsker's Theorem, converges weakly to a standard
Brownian motion in the continuous time limit as $h\rightarrow0$ (see
Billingsley \cite[Thm. 14.1 and Example 12.3]{billingsley}, for example).

\medskip

\noindent Scenario 2: You are told again that $Y=0$ for every urn. However,
you are given less information than previously about the other colors.
Specifically, you are told that for each urn the proportion of $R$ lies in the
interval $\left[  \frac{1}{2}-\frac{1}{2}\kappa h^{1/2},\frac{1}{2}+\frac
{1}{2}\kappa h^{1/2}\right]  $, for some fixed $\kappa>0$. Thus the
composition of the urn at any time $t$ could be given by a measure of the form
$p^{\mu_{t}}=\left(  \frac{1}{2}+\frac{1}{2}\mu_{t}h^{1/2},\frac{1}{2}%
-\frac{1}{2}\mu_{t}h^{1/2},0\right)  $, for some $\mu_{t}$ satisfying $\mid
\mu_{t}\mid\leq\kappa$. The increment $dB_{t}$ has mean and variance under
$p^{\mu_{t}}$ given by\footnote{$o\left(  h\right)  $ represents a function
such that $o\left(  h\right)  /h\rightarrow0$ as $h\rightarrow0$.}%
\[
E\left(  dB_{t}\right)  =\mu_{t}h\text{ ~and }var\left(  dB_{t}\right)
=h-(\mu_{t}h)^{2}=h+o\left(  h\right)  \text{.}%
\]
Accordingly, the weaker information about the composition of urns implies
ambiguity about drift per unit time, but up to the $o\left(  h\right)  $
approximation, it does not affect the corresponding one-step-ahead variance.

The preceding is the building block of the Chen and Epstein (2002) continuous
time model of ambiguity about drift.\footnote{It corresponds to the special
case of their model called $\kappa$-ignorance.} The transition from discrete
to continuous time amounts to a minor variation of the convergence result
noted for Scenario 1 (see Skiadas \cite{skiadas2011} for some details). The
sets $\{p^{\mu_{t}}:\mid\mu_{t}\mid\leq\kappa\}$, $t=0,h,2h,...,\left(
n-1\right)  h,nh$, of one-step-ahead measures can be combined to construct a
set $\mathcal{P}$ of priors over the set $\Omega$ of possible trajectories for
$B$. It is not difficult to see that the priors are mutually \emph{equivalent}%
, that is, they all agree on which events are null (have zero probability).
Further, as described in Section \ref{section-recursive}, equivalence holds in
the continuous time limit. Consequently, the model with ambiguous drift can be
formulated within a probability space framework with ambient probability
measure $P_{0}$ according to which $B$ is a standard Brownian motion.
Alternative hypotheses about the true probability law can be expressed via
densities with respect to $P_{0}$.

\medskip

\noindent Scenario 3: Turn now to a model having ambiguity only about
volatility. You are told that $R=G$, thus eliminating uncertainty about the
relative composition of $R$ versus $G$. However, the information about $Y$ is
weakened and you are told only that $Y\leq20$.\footnote{This scenario is
adapted from Levy et al. \cite{levy}.}

Any probability measure over trajectories consistent with these facts makes
$B$ a martingale. In that sense, there is certainty that $B$ is a martingale.
However, the one-step-ahead variance $\sigma_{t}^{2}h$ depends on the number
of yellow balls and thus is ambiguous - it equals $p_{t}h$, where we know only
that $0\leq1-p_{t}\leq0.2$, or%
\begin{equation}
.8=\underline{\sigma}^{2}\leq\sigma_{t}^{2}\leq\overline{\sigma}^{2}%
=1\text{.}\label{sigmabound}%
\end{equation}

\noindent Because urns are perceived to be independent, they may differ in
actual composition. Therefore, any value for $\sigma_{t}$ in the interval
$\left[  \underline{\sigma},\overline{\sigma}\right]  $ could apply at any
time. Independence implies also that past draws do not reveal anything about
the future and ambiguity is left undiminished. This is an extreme case that is
a feature of this example and is a counterpart of the assumption of i.i.d.
increments in the binomial tree.

By a generalization of Donsker's Theorem (Yuan \cite{yuan}), the above
trinomial model converges weakly (or \textquotedblleft in
distribution\textquotedblright) to a continuous time model on the interval
$\left[  0,T\right]  $ as the time period length $h$ goes to $0$.\footnote{To
clarify the meaning of weak convergence, consider the set $\Omega$ of
continuous trajectories on $[0,T]$ that begin at the origin. For each period
length $h$, identify any discrete time trajectory with a continuous path on
$\left[  0,T\right]  $ obtained by linear interpolation. Then each hypothesis
about the compositions of all urns implies a probability measure on $\Omega$.
By varying over all hypotheses consistent with the above description of the
urns, one obtains a set $\mathcal{P}^{h}$ of probability laws on $\Omega$.
Suppose that $\mathcal{P}$ is a given set of measures on $\Omega$. Then say
that $\mathcal{P}^{h}$ converges weakly to $\mathcal{P}$ if $\sup
_{P\in\mathcal{P}^{h}}E^{P}f$ converges to $\sup_{P\in\mathcal{P}}E^{P}f$ for
every function $f:\Omega\rightarrow\mathbb{R}$ that is bounded and suitably
continuous. The cited result by Yuan implies this convergence for the set
$\mathcal{P}$ constructed below corresponding to the special case of our model
for which volatility is constrained by (\ref{sigmabound}). See also Dolinsky
et al. \cite{dolinsky} for a related result.}

\bigskip

The limiting continuous time model inherits from the discrete time trinomial
the interpretation that it models certainty that the driving process
$B=(B_{t})$ is a martingale, whereas volatility is known only up to the
interval $\left[  \underline{\sigma},\overline{\sigma}\right]  $. To be more
precise about the meaning of volatility, let the quadratic variation process
of $B$ be defined by%
\begin{equation}
\langle B\rangle_{t}(\omega)=\underset{\bigtriangleup t_{k}\rightarrow0}{\lim
}~\underset{t_{k}\leq t}{\Sigma}\mid B_{t_{k+1}}(\omega)-B_{t_{k}}(\omega
)\mid^{2}\label{Blim}%
\end{equation}
where $0=t_{1}<\ldots<t_{n}=t$ and $\bigtriangleup t_{k}=t_{k+1}-t_{k}%
$.\footnote{By Follmer \cite{follmer} and Karandikar \cite{ka}, the above
limit exists almost surely for every measure that makes $B$ a martingale.
Because there is certainty that $B$ is a martingale, this limited universality
is all we need.} Then the volatility $\left(  \sigma_{t}\right)  $ of $B$ is
defined by
\[
d\langle B\rangle_{t}=\sigma_{t}^{2}dt\text{.}%
\]
Therefore, the interval constraint on volatility can be written also in the
form
\begin{equation}
\underline{\sigma}^{2}t\leq\langle B\rangle_{t}\leq\overline{\sigma}%
^{2}t\text{.}\label{sigma-ineq}%
\end{equation}

The preceding defines the stochastic environment. Consumption and other
processes are defined accordingly (for example, they are required to be
adapted to the natural filtration generated by $B$). We emphasize that our
model is much more general than suggested by this outline. Importantly, the
interval $\left[  \underline{\sigma},\overline{\sigma}\right]  $ can be time
and state varying, and the dependence on history of the interval at time $t$
is unrestricted, thus \emph{permitting any model of how ambiguity varies with
observation (that is, learning) to be accommodated}. In addition, we admit
multidimensional driving processes and also ambiguity about both drift and volatility.

As noted earlier, ambiguity about volatility leads to a set of
\emph{nonequivalent} priors, that is, to disagreement between priors as to
what events are possible (or to ambiguity about what is possible). To see
this, let $B$ be a Brownian motion under $P_{0}$ and denote by
$P^{\underline{\sigma}}$ and $P^{\overline{\sigma}}$ the probability
distributions over continuous paths induced by the two processes
$($\underline{$\sigma$}$B_{t}) $ and $\left(  \overline{\sigma}B_{t}\right)
$. Given the ambiguity described by (\ref{sigma-ineq}), $P^{\underline{\sigma
}}$ and $P^{\overline{\sigma}}$ are two alternative hypotheses about the
probability law driving uncertainty. It is apparent that they are mutually
singular, and hence not equivalent, because\footnote{Two measures $P$ and
$P^{\prime}$ on $\Omega$ are singular if there exists $A\subset\Omega$ such
that $P\left(  A\right)  =1$ and $P^{\prime}\left(  A\right)  =0$. They are
equivalent, if for every $A$, $P\left(  A\right)  =0$ if and only if
$P^{\prime}\left(  A\right)  =0$. Thus $P$ and $P^{\prime}$ singular implies
that they are not equivalent, but the converse is false.}
\begin{equation}
P^{\underline{\mathbf{\sigma}}}(\{\langle B\rangle_{T}=\underline{\sigma}%
^{2}T\})=1=P^{\overline{\mathbf{\sigma}}}(\{\langle B\rangle_{T}%
=\overline{\sigma}^{2}T\})\text{.}\label{Psingular}%
\end{equation}

We caution against a possible conceptual misinterpretation of (\ref{Psingular}%
). If $P^{\underline{\sigma}}$ and $P^{\overline{\sigma}}$ were the only two
hypotheses being considered, then ambiguity could be eliminated quickly
because one can approximate volatility locally as in (\ref{Blim}) and thus use
observations on a short time interval to differentiate between the two
hypotheses. This is possible because of the tight connection between past and
future volatility imposed in each of $P^{\underline{\sigma}}$ and
$P^{\overline{\sigma}}$. As discussed earlier, this is not the kind of
ambiguity we have in mind. The point of (\ref{Psingular}) is only to
illustrate nonequivalence as simply as possible. Importantly, such
nonequivalence of priors is a feature also of the more complex and interesting
cases at which the model is directed.

An objection to modeling ambiguity about possibility might take the form:
\textquotedblleft If distinct priors (or models) are not equivalent, then one
can discriminate between them readily. Therefore, when studying nontransient
ambiguity there is no loss in restricting priors to be
equivalent.\textquotedblright\ The connection between past and future is again
the core issue (as in the preceding subsection). Consider the individual at
time $t$ and her beliefs about the future. The source of ambiguity is her
concern with locally time varying and poorly understood factors. This limits
her confidence in predictions about the immediate future, or `next step', to a
degree that depends on history but that is not eliminated by the retrospective
empirical discrimination between models. At a formal level, Epstein and
Schneider \cite{es2003} show that when backward induction reasoning is added
to multiple priors utility, then the individual behaves \emph{as if} the set
of conditionals entertained at any time $t$ and state does not vary with
marginal prior beliefs on time $t$ measurable uncertainty. (They call this
property rectangularity.) Thus looking back on past observations at $t$, even
though the individual might be able to dismiss some priors or models as being
inconsistent with the past, this is unimportant for prediction because the set
of conditional beliefs about the future is unaffected.

\section{Utility\label{section-utility}}

Many components of the formal setup are typical in continuous time asset
pricing. Time $t$ varies over the finite horizon $[0,T]$. Paths or
trajectories of the driving process are assumed to be continuous and thus are
modeled by elements of $C^{d}([0,T])$, the set of all $\mathbb{R}^{d}$-valued
continuous functions on $[0,T]$, endowed with the sup norm. The generic path
is $\omega=(\omega_{t})_{t\in\lbrack0,T]}$, where we write $\omega_{t}$
instead of $\omega\left(  t\right)  $. All relevant paths begin at $0$ and
thus we define the canonical state space to be
\[
\Omega=\left\{  \omega=\left(  \omega_{t}\right)  \in C^{d}([0,T]):\omega
_{0}=0\right\}  \text{.}%
\]

The coordinate process $\left(  B_{t}\right)  $, where $B_{t}(\omega
)=\omega_{t}$,\ is denoted by $B$. Information is modeled by the filtration
$\mathcal{F}=\{\mathcal{F}_{t}\}$ generated by $B$. Let $P_{0}$ be the Wiener
measure on $\Omega$ so that $B$ is a Brownian motion under $P_{0}$.

Consumption processes $c$ take values in $C$, a convex subset of
$\mathbb{R}^{\ell}$. The objective is to formulate a suitable utility function
on a domain $D$ of consumption processes.

\subsection{Recursive Utility under Equivalence\label{section-recursive}}

For perspective, we begin by outlining the Chen-Epstein model where there is
ambiguity only about drift. This is the continuous time counterpart of
Scenario 2 in Section \ref{section-trinomial}.

If $P_{0}$ describes the individual's beliefs, then following Duffie and
Epstein \cite{DE} utility may be defined by:\footnote{Below we often suppress
$c$ and write $V_{t}$ instead of $V_{t}\left(  c\right)  $. The dependence on
the state $\omega$ is also frequently suppressed.}%

\begin{equation}
V_{t}^{P_{0}}\left(  c\right)  =E^{P_{0}}[\int_{t}^{T}f(c_{s},V_{s}^{P_{0}%
})ds\mid\mathcal{F}_{t}],\;0\leq t\leq T.\label{SDU}%
\end{equation}
Here $V_{t}^{P_{0}}$ gives the utility of the continuation $(c_{s})_{s\geq t}
$ and $V_{0}^{P_{0}}$ is the utility of the entire process $c$. The function
$f$ is a primitive of the specification, called an \emph{aggregator}. The most
commonly used aggregator has the form%
\begin{equation}
f\left(  c_{t},v\right)  =u\left(  c_{t}\right)  -\beta v\text{, ~\ }\beta
\geq0\text{,}\label{fstandard}%
\end{equation}
which delivers the expected utility specification
\begin{equation}
V_{t}\,\left(  c\right)  =\,E^{P_{0}}\left[  \int_{t}^{T}\,e^{-\beta
(s-t)}\,u(c_{s})\,ds\mid\,\mathcal{F}_{t}\right]  \text{.}\label{eu}%
\end{equation}
The use of more general aggregators permits a partial separation of risk
aversion from intertemporal substitution.

To admit a concern with model uncertainty, Chen and Epstein replace the single
measure $P_{0}$ by a set $\mathcal{P}^{\Theta}$ of measures equivalent to
$P_{0}$. This is done by specifying a suitable set of densities. For each
well-behaved $\mathbb{R}^{d}$-valued process $\theta=(\theta_{t})$, called a
density generator, let%
\[
z_{t}^{\theta}\,\equiv\,\exp\left\{  -\tfrac{1}{2}\,\int_{0}^{t}\mid\theta
_{s}\mid^{2}ds\,-\,\int_{0}^{t}\,\theta_{s}^{\top}\,dB_{s}\right\}  \text{,
\thinspace}0\leq t\leq T\text{,}%
\]
and let $P^{\theta}$ be the probability measure on $(\Omega,\mathcal{F})$ with
density $\,z_{T}^{\theta}$, that is,
\begin{equation}
\frac{dP^{\theta}}{dP_{0}}\,=\,z_{T}^{\theta}\text{; \thinspace more
generally, }\left.  \frac{dP^{\theta}}{dP}\right\vert _{\mathcal{F}_{t}%
}\,=\,z_{t}^{\theta}\,\,\,\text{for each }t\text{.}\label{Qtheta}%
\end{equation}
\thinspace\thinspace Given a set $\Theta$ of density generators, the
corresponding set of priors is
\begin{equation}
\mathcal{P}^{\Theta}\,=\,\{\,P^{\theta}\,:\,\theta\in\Theta\text{ and
}P^{\theta}\text{ is defined by (\ref{Qtheta})\thinspace}\}\text{.}%
\label{PthetaCE}%
\end{equation}
By construction, all measures in $\mathcal{P}^{\Theta}$ are equivalent to
$P_{0}$. Because the role of $P_{0}$ is only to define null events, any other
member of $\mathcal{P}^{\Theta}$ could equally well serve as the reference measure.

Continuation utilities are defined by:%

\begin{equation}
V_{t}=\underset{P\in\mathcal{P}^{\Theta}}{\inf}E^{P}[\int_{t}^{T}f(c_{s}%
,V_{s})ds\mid\mathcal{F}_{t}]\text{.}\label{envelope}%
\end{equation}
An important property of the utility process is dynamic consistency, which
follows from the following recursivity: For every $c$ in $D$,
\begin{equation}
V_{t}=\,\min_{P\in\mathcal{P}^{\Theta}}E_{P}\left[  \int_{t}^{\tau}%
\,f(c_{s},V_{s})\,ds\,+\,V_{\tau}\,\mid\,\mathcal{F}_{t}\right]  \text{,
\thinspace\thinspace}0\leq t<\tau\leq T\text{.}\label{rc}%
\end{equation}
Regarding interpretation, by the Girsanov theorem $B_{t}+\int_{0}^{t}%
\theta_{s}ds$ is a Brownian motion under $P^{\theta}$. Thus as $\theta$ varies
over $\Theta$ and $P^{\theta}$ varies over $\mathcal{P}^{\Theta}$, alternative
hypotheses about the drift of the driving process are defined. Accordingly,
the infimum suggests that the utility functions $V_{t}$ exhibit an aversion to
ambiguity about the drift. Because $B_{t}$ has variance-covariance matrix
equal to the identity according to all measures in $\mathcal{P}^{\Theta}$,
there is no ambiguity about volatility. Neither is there any uncertainty about
what is possible because $P_{0}$ defines which events are null.

\bigskip

There is a limited sense in which the preceding framework is adequate for
modeling also ambiguity about volatility. For example, suppose that the
driving process is $\left(  X_{t}\right)  $, where $dX_{t}=\sigma_{t}dB_{t}$,
(where $B$ is a Brownian motion under $P_{0}$ and) where the volatility is
thought to evolve according to%
\[
d\sigma_{t}=\theta_{t}dt+v_{t}dB_{t}\text{.}%
\]
Here the drift $(\theta_{t})$ is ambiguous in the above sense, and the
volatility of volatility $(v_{t})$ is a fixed stochastic process, for example,
it might be constant as in many stochastic volatility models. Thus the
difficulty of finding a specification for $\left(  \sigma_{t}\right)  $ in
which one can have complete confidence is moved one level from volatility to
its volatility. This constitutes progress if there is greater evidence about
vol of vol and if model implications are less sensitive to misspecifications
of the latter. We suspect that in many modeling situations neither is true.
Moreover, this approach cannot intermediate between, or robustify, the
stochastic volatility models that have been used in the empirical literature
(Section \ref{section-why}).

\subsection{The set of priors\label{section-priors}}

The objective is to specify beliefs, in the form of a set of priors
generalizing (\ref{PthetaCE}), that captures ambiguity about both drift and
volatility. Another key ingredient is conditioning. The nonequivalence of
priors (illustrated by (\ref{Psingular})) poses a particular difficulty for
updating because of the need to update beliefs conditional on events having
zero probability according to some, but not all, priors. Once these steps are
completed, continuation utilities can be defined (apart from technical
details) as in (\ref{envelope}); see the next section.

The construction of the set of priors can be understood by referring back to
the binomial and trinomial examples in the introduction. In Scenario 2, the
composition of all urns is specified by fixing $\mu=\left(  \mu_{t}\right)  $
with $\mid\mu_{t}\mid\leq\kappa$. Define $X^{\mu}=\left(  X_{t}^{\mu}\right)
$ by%
\[
dX_{t}^{\mu}=\mu_{t}h+dB_{t}\text{, \ }X_{0}^{\mu}=0\text{.}%
\]
Then $X^{\mu}$ and $P_{0}$ induce a distribution $P^{\mu}$ over trajectories;
and as one varies over all choices of $\mu$, one obtains the set of priors
$\mathcal{P}$ described earlier. Thus beliefs are described indirectly through
the set $\{\mu=\left(  \mu_{t}\right)  :\mid\mu_{t}\mid\leq\kappa\}$ of
alternative hypotheses about the drift of the driving process. Similarly in
Scenario 3, where the composition of all urns is specified by by fixing
$\sigma=\left(  \sigma_{t}\right)  $ with $\underline{\sigma}\leq\sigma
_{t}\leq\overline{\sigma}$. If $X^{\sigma}=\left(  X_{t}^{\sigma}\right)  $ is
defined by%
\[
dX_{t}^{\sigma}=\sigma_{t}dB_{t}\text{, \ }X_{0}^{\sigma}=0\text{,}%
\]
then $X^{\sigma}$ and $P_{0}$ induce a distribution $P^{\sigma}$ over
trajectories; and as one varies over all choices of $\sigma$, one obtains the
set of priors $\mathcal{P}$ described earlier. Thus beliefs are described
indirectly through a set of alternative hypotheses about the volatility of the
driving process.

The preceding construction is readily generalized to permit a vector-valued
driving process ($d\geq1$), ambiguity about both drift and volatility, and to
allow ambiguity at any time $t$ to depend on history. \ We describe the
corresponding construction in continuous time.\footnote{The reader is referred
to Epstein and Ji \cite{eji} for a general and mathematically rigorous
development.}

The individual is not certain that the driving process has zero drift and/or
unit variance (where $d=1$). Accordingly, she entertains a range of
alternative hypotheses $X^{\theta}=(X_{t}^{\theta})$ parametrized by $\theta$.
Here $\theta_{t}=(\mu_{t},\sigma_{t})$ is an $\mathcal{F}$-progressively
measurable process with values in $\mathbb{R}^{d}\times\mathbb{R}^{d\times d}$
that describes a conceivable process for drift $\mu=\left(  \mu_{t}\right)  $
and for volatility $\sigma=\left(  \sigma_{t}\right)  $.\footnote{Write
$\theta=(\mu,\sigma)$.} Available information leads to the constraint on drift
and volatility pairs given by%
\begin{equation}
\theta_{t}\left(  \omega\right)  \in\Theta_{t}\left(  \omega\right)  \text{,
\ for all }\left(  t,\omega\right)  \text{,}\label{THETAt}%
\end{equation}
where $\Theta_{t}\left(  \omega\right)  $ is a subset of $\mathbb{R}^{d}%
\times\mathbb{R}^{d\times d}$. The $\Theta_{t}$'s are primitives of the
model.\footnote{See our companion paper for the technical regularity
conditions assumed for $\left(  \Theta_{t}\right)  $.} In the trinomial model
expanded in the obvious way to include also ambiguity about drift,
\[
\Theta_{t}\left(  \omega\right)  =\left[  -\kappa,\kappa\right]  \times
\lbrack\underline{\sigma},\overline{\sigma}]\text{, ~for all }\left(
t,\omega\right)  \text{.}%
\]

In general, the dependence of $\Theta_{t}\left(  \omega\right)  $ on the
history corresponding to state $\omega$ permits the model to accommodate
learning. Moreover, because the form of history dependence is unrestricted
(apart from technical regularity conditions), so is the nature of learning.
Just as in the Chen-Epstein model, we provide a framework within which
additional structure modeling learning can be added.

Two other examples might be helpful. The robust stochastic volatility model
described in the introduction corresponds to taking
\[
\Theta_{t}\left(  \omega\right)  =\{0\}\times\lbrack\underline{\sigma}%
_{t}\left(  \omega\right)  ,\overline{\sigma}_{t}\left(  \omega\right)
]\text{, }%
\]
where $\underline{\sigma}_{t}\left(  \omega\right)  $ and $\overline{\sigma
}_{t}\left(  \omega\right)  $ are given by (\ref{robustvol}). When $d>1$, one
way to robustify is through the restriction%
\[
\Theta_{t}\left(  \omega\right)  =\left\{  \sigma\in\mathbb{R}^{d\times
d}:\sigma_{t}^{1}\left(  \omega\right)  \left(  \sigma_{t}^{1}\left(
\omega\right)  \right)  ^{\intercal}\leq\sigma\sigma^{\intercal}\leq\sigma
_{t}^{2}\left(  \omega\right)  \left(  \sigma_{t}^{2}\left(  \omega\right)
\right)  ^{\intercal}\right\}  \text{,}%
\]
though other natural specifications exist in the multidimensional case.

The model is flexible in the way it relates ambiguity about drift and
ambiguity about volatility. Thus, as a final example, suppose that drift and
volatility are thought to move together. Then joint ambiguity is captured by
specifying%
\begin{equation}
\Theta_{t}(\omega)=\{(\mu,\sigma)\in\mathbb{R}^{2}:\mu=\mu_{\min}%
+z,~\sigma^{2}=\sigma_{\min}^{2}+2z/\gamma,~0\leq z\leq\overline{z}_{t}\left(
\omega\right)  \}\text{,}\label{joint}%
\end{equation}
where $\mu_{\min}$, $\sigma_{\min}^{2}$ and $\gamma>0$ are fixed and known
parameters.\footnote{This specification is adapted from Epstein and Schneider
\cite{esAR}.}

Given a hypothesis $\theta$ about drift and volatility, the implication for
the driving process is that it is given by the unique solution $X^{\theta
}=\left(  X_{t}^{\theta}\right)  $ to the following stochastic differential
equation (SDE) under $P_{0}$:%
\begin{equation}
dX_{t}^{\theta}=\mu_{t}(X_{\cdot}^{\theta})dt+\sigma_{t}(X_{\cdot}^{\theta
})dB_{t}\text{,}\;\ \ X_{0}^{\theta}=0\text{,}\;t\in\lbrack0,T]\text{.}%
\label{Xtheta}%
\end{equation}
We restrict the process $\theta$ further so that a unique strong solution
$X^{\theta}$ to the SDE exists. Denote by $\Theta$ the set of all processes
$\theta$ satisfying the latter and also (\ref{THETAt}).

As in the discrete time examples, $X^{\theta}$ and $P_{0}$ induce a
probability measure $P^{\theta}$ on $(\Omega,\mathcal{F}_{T})$:%
\begin{equation}
P^{\theta}(A)=P_{0}(\{\omega:X^{\theta}(\omega)\in A\})\text{, }%
A\in\mathcal{F}_{T}\text{.}\label{PthetaP0}%
\end{equation}
Therefore, we arrive at the set $\mathcal{P}^{\Theta}$ of priors on the set of
continuous trajectories given by
\begin{equation}
\mathcal{P}^{\Theta}\mathcal{=\{}P^{\theta}:\theta\in\Theta\}.\label{Ptheta}%
\end{equation}
Fix $\Theta$ and denote the set of priors simply by $\mathcal{P}$. This set of
priors is used, as in the Gilboa-Schmeidler model, to define utility and guide
choice between consumption processes.

\begin{remark}
\label{remark-recap}Here is a recap. The set $\mathcal{P}$ consists of priors
over $\Omega$, the space of continuous trajectories for the driving process.
$B$ denotes the coordinate process, $B_{t}\left(  \omega\right)  =\omega_{t}$.
It is a Brownian motion under $P_{0}$, which may or may not lie in
$\mathcal{P}$, but $B$ is typically not a Brownian motion relative to (other)
priors in $\mathcal{P}$. Indeed, different priors $P$ typically imply
different conditional expectations $E_{t}^{P}\left(  B_{t+\bigtriangleup
t}-B_{t}\right)  $ and $E_{t}^{P}\left(  \langle B\rangle_{t+\bigtriangleup
t}-\langle B\rangle_{t}\right)  $.{\LARGE \ }This is the justification for
interpreting $\mathcal{P}$ as modeling ambiguity about the drift and
volatility of the driving process. The preceding should be viewed as one way
to construct $\mathcal{P}$, but not necessarily as a description of the
individual's thought processes. The model's objective is to describe behavior
that can be thought of \textquotedblleft as if\textquotedblright\ being
derived from a maxmin objective function using the above set of priors.
\end{remark}

The construction of utility requires that first we show how beliefs, through
the set $\mathcal{P}$, lead to natural definitions of \textquotedblleft
expectation\textquotedblright\ and \textquotedblleft conditional
expectation.\textquotedblright\ The former is straightforward. For any random
variable $\xi$\ on $(\Omega,\mathcal{F}_{T})$, if $\sup_{P\in\mathcal{P}}%
E_{P}\xi<\infty$, define its (nonlinear) expectation by%
\begin{equation}
\hat{E}\xi=\sup_{P\in\mathcal{P}}E_{P}\xi.\label{Ehat}%
\end{equation}
Because we will assume that the individual is concerned with worst-case
scenarios, below we use the fact that%
\[
\inf_{P\in\mathcal{P}}E_{P}\xi=-\hat{E}[-\xi]\text{.}%
\]

Conditional beliefs and expectation are not as clear cut because of the need,
mentioned above, to update beliefs conditional on events having zero
probability according to some priors. A naive approach to defining conditional
expectation would be to use the standard conditional expectation $E_{P}%
[\xi\mid\mathcal{F}_{t}]$ for each $P$ in $\mathcal{P}$ and then to take the
(essential) supremum over $\mathcal{P}$. Such an approach immediately
encounters a roadblock due to the nonequivalence of priors. The conditional
expectation $E_{P}[\xi\mid\mathcal{F}_{t}]$ is well defined only $P$-almost
surely, while speaking informally, conditional beliefs and expectation must be
defined at every node deemed possible by some prior in $\mathcal{P}$. The
economic rationale is simple. Suppose that $P$ and $P^{\prime}$ are two
nonequivalent priors held (for example) at time $0$, and consider updating at
$t>0$. Let $A$ be an event, measurable at time $t$, such that $P\left(
A\right)  =0<P^{\prime}\left(  A\right)  $. Then $A$ is conceivable according
to the individual's ex ante perception. Consequently, beliefs at time $t$
conditional on $A$ are relevant both for ex post choice and also for ex ante
choice, for example, if the individual reasons by backward induction.
Therefore, the ex ante perception represented by $P$ should also be updated
there, even though $A$ was deemed impossible ex ante according to $P$.

This difficulty can be overcome because for every admissible hypothesis
$\theta$, $\theta_{t}\left(  \omega\right)  $ is defined for \emph{every}
$\left(  t,\omega\right)  $, that is, the primitives specify a hypothesized
instantaneous drift-volatility pair everywhere in the tree. This feature of
the model resembles the approach adopted in the theory of extensive form
games, namely the use of conditional probability systems, whereby conditional
beliefs at \emph{every} node are specified as primitives, obviating the need
to update.\footnote{It resembles also the approach in the discrete time model
in Epstein and Schneider \cite{es2003}, where roughly, conditional beliefs
about the next instant for every time and history are adopted as primitives
and are pasted together by backward induction to deliver the ex ante set of
priors.} We show that a solution to the updating problem is also available
here (though it requires nontrivial mathematical arguments - see Epstein and
Ji \cite{eji} for details, rigorous statements and supporting proofs).

To outline it, let $\theta=\left(  \theta_{s}\right)  $ be a conceivable
scenario ex ante and fix a node $\left(  t,\omega\right)  $. By definition of
$\theta$, the continuation of $\theta$ is seen by the individual \emph{ex
ante} as a conceivable continuation from time $t$ along the history $\omega$.
We assume that then it is also seen as a conceivable scenario \emph{ex post}
conditionally on $\left(  t,\omega\right)  $, thus ruling out surprises or
unanticipated changes in outlook. Then, paralleling (\ref{Xtheta}), each such
conditional scenario has an implication for the driving process conditionally
on $\left(  t,\omega\right)  $. The implied process and $P_{0}$ induce a
probability measure $P_{t}^{\theta,\omega}$ on $\Omega$, denoted simply by
$P_{t}^{\omega}$ with $\theta$ suppressed when it is understood that
$P=P^{\theta}$. The crucial facts are that, for each $P$ in $\mathcal{P}$, (i)
$P_{t}^{\omega}$ is defined for every $t$ and $\omega$, and (ii)
$P_{t}^{\omega}$ is a version of the regular $\mathcal{F}_{t}$-conditional
probability of $P$.\footnote{For any probability measure $P$ on the canonical
space $\Omega$, a corresponding \emph{regular }$\mathcal{F}_{t}$%
-\emph{conditional probability }$P_{t}^{\omega}$ is defined to be any mapping
$P_{t}^{\omega}:\Omega\times\mathcal{F}_{T}\rightarrow\lbrack0,1]$ satisfying
the following conditions: (i) for any $\omega$, \ $P_{t}^{\omega}$ is a
probability measure on $(\Omega,\mathcal{F}_{T})$. (ii) for any $A\in F_{T}$,
$\omega\rightarrow P_{t}^{\omega}(A)$\ is $\mathcal{F}_{t}$-measurable. (iii)
for any $A\in\mathcal{F}_{T}$, $E^{P}[1_{A}\mid\mathcal{F}_{t}](\omega
)=P_{t}^{\omega}(A),\;P$-$a.s.$} The set of all such conditionals obtained as
$\theta$ varies over $\Theta$ is denoted $\mathcal{P}_{t}^{\omega}$, that is,
\begin{equation}
\mathcal{P}_{t}^{\omega}=\left\{  P_{t}^{\omega}:P\in\mathcal{P}\right\}
\text{.}\label{regcond}%
\end{equation}
We take $\mathcal{P}_{t}^{\omega}$ to be the individual's set of priors
conditional on $\left(  t,\omega\right)  $. Then, the conditional expectation
of suitable random variables $\xi$ is defined by%
\[
\hat{E}[\xi\mid\mathcal{F}_{t}]\left(  \omega\right)  =\sup_{P\in
\mathcal{P}_{t}^{\omega}}E_{P}\xi\text{, \ for every }(t,\omega)\text{.}%
\]
This completes the prerequisites regarding beliefs for defining utility.

\subsection{The definition of utility}

Let $D$ be a domain of consumption processes and defer elaboration until the
next section. For each $c$ in $D$, define its continuation utility
$V_{t}\left(  c\right)  $, or simply $V_{t}$, by%

\begin{equation}%
\begin{array}
[c]{cl}%
V_{t} & =-\hat{E}[-\int_{t}^{T}f(c_{s},V_{s})ds\mid\mathcal{F}_{t}]
\end{array}
\text{.}\label{Vt}%
\end{equation}
This definition parallels the Chen and Epstein definition (\ref{envelope}). In
particular, $f$ is an aggregator that is assumed to satisfy suitable
measurability, Lipschitz and integrability conditions. Under these conditions
and for a suitable domain $D$, there is a unique utility process $\left(
V_{t}\left(  c\right)  \right)  $ solving (\ref{Vt}) for each $c$ in $D$, that
is, utility is well-defined.

For the standard aggregator (\ref{fstandard}), utility admits the closed-form
expression%
\begin{equation}
V_{t}\left(  c\right)  =-\hat{E}[-\int_{t}^{T}u(c_{s})e^{-\beta s}%
ds\mid\mathcal{F}_{t}].\label{Vstandard}%
\end{equation}
More generally, closed-from expressions are rare.

The following example illustrates the effect of volatility ambiguity.

\begin{example}
[Closed form]\label{example-closedform}Suppose that there is no ambiguity
about the drift, and that ambiguity about volatility is captured by the fixed
interval $\left[  \underline{\sigma},\overline{\sigma}\right]  \subset
\mathbb{R}_{++} $. Consider consumption processes that are certain and
constant, at level $0$ for example, on the time interval $[0,1)$, and that
yield constant consumption on $[1,T]$ at a level that depends on the state
$\omega_{1}$ at time 1. Specifically, let
\[
c_{t}(\omega)=\psi(\omega_{1})\text{, \ for }1\leq t\leq T\text{,}%
\]
where $\psi:\mathbb{R}^{1}\rightarrow\mathbb{R}_{+}^{1}$. For simplicity,
suppose further that $u$ is linear. Then time $0$ utility evaluated using
(\ref{Vstandard}), is, ignoring irrelevant constants,%
\[
V_{0}=-\hat{E}[-\psi(\omega_{1})].
\]
If $\psi$ is a convex function, then\footnote{See Levy et al. \cite{levy} and
Peng \cite{P-2010}.}
\[
V_{0}\left(  c\right)  =\frac{1}{\sqrt{2\pi}}\int\nolimits_{-\infty}^{\infty
}\psi(\underline{\sigma}^{2}y)\exp(-\tfrac{y^{2}}{2})dy\text{,}%
\]
and if $\psi$ is concave, then
\begin{equation}
V_{0}\left(  c\right)  =\frac{1}{\sqrt{2\pi}}\int\nolimits_{-\infty}^{\infty
}\psi(\bar{\sigma}^{2}y)\exp(-\tfrac{y^{2}}{2})dy.\label{concave}%
\end{equation}
There is an intuitive interpretation for these formulae. Given risk
neutrality, the individual cares only about the expected value of consumption
at time $1$. The issue is expectation according to which probability law? For
simplicity, consider the following concrete specifications:%
\[
\psi_{1}\left(  x\right)  =\mid x-\kappa\mid\text{, and }\psi_{2}\left(
x\right)  =-\mid x-\kappa\mid\text{.}%
\]
Then $\psi_{1}$ is convex and $\psi_{2}$ is concave. If we think of the the
driving process as the price of a stock, then $\psi_{1}(\cdot)$ can be
interpreted as a straddle - the sum of a European put and a European call
option on the stock at the common strike price $\kappa$ and expiration date
$1$. (We are ignoring nonnegativity constraints.) A straddle pays off if the
stock price moves, whether up or down, and thus constitutes a bet on
volatility. Accordingly, the worst case scenario is that the price process has
the lowest possible volatility $\underline{\sigma}$. In that case, $\omega
_{1}$ is $N\left(  0,\underline{\sigma}^{2}\right)  $ and the indicated
expected value of consumption follows. Similarly, $\psi_{2}\left(
\cdot\right)  $ describes the corresponding short position and amounts to a
bet against volatility. Therefore, the relevant volatility for its worst case
evaluation is the largest possible value $\overline{\sigma}$, consistent with
the expression for utility given above.

When the function $\psi$ is neither concave nor convex globally, closed-form
expressions for utility are available only in extremely special and
unrevealing cases. However, a generalization to $d$-dimensional processes,
$d\geq1$, is available and will be used below. Let there be certainty that the
driving process is a martingale and let the volatility matrix $\sigma_{t}$ in
(\ref{Xtheta}) be restricted to lie in the compact and convex set
$\Gamma\subset\mathbb{R}^{d\times d}$ such that, for all $\sigma$ in $\Gamma$,
$\sigma\sigma^{\top}\geq\hat{a}$ for some positive definite matrix
$\widehat{a}$. Consumption is as above except that, for some $a\in
\mathbb{R}^{d}$,
\[
c_{t}(\omega)=\psi(a^{\top}\omega_{1})\text{, \ for }1\leq t\leq T\text{.}%
\]
Let \underline{$\sigma$} be any solution to $\min_{\sigma\in\Gamma}tr\left(
\sigma\sigma^{\top}aa^{\top}\right)  $ and let $\overline{\sigma}$ be any
solution to \newline$\max_{\sigma\in\Gamma}tr\left(  \sigma\sigma^{\top
}aa^{\top}\right)  $. If $\psi$ is convex (concave), then the worst-case
scenario is that $\sigma_{t}=$\underline{$\sigma$} ($\overline{\sigma}$) for
all $t$. Closed-form expressions for utility follow immediately.
\end{example}

The domain $D$ of consumption processes, and the ambient space containing
utility processes, are defined precisely in our companion paper (see also the
next section). Here we mention briefly a feature of these definitions that
reveals a great deal about the nature of formal analysis when priors are not
equivalent. When the ambient framework is a probability space $\left(
\Omega,P_{0}\right)  $, and thus $P_{0}$ is used to define null events, then
random variables and stochastic processes are defined only up to the $P_{0}%
$-almost sure qualification. Thus $P_{0}$ is an essential part of the
definition of all formal domains. However, ambiguity about volatility leads to
a set $\mathcal{P}$ of priors that do not agree about which events have zero
probability. Therefore, we follow Denis and Martini \cite{DM} and define
appropriate domains of stochastic processes by using the entire set of
probability measures $\mathcal{P}$. Accordingly, two consumption processes
$c^{\prime}$ and $c$ are identified, and we write $c^{\prime}=c$, if for every
$t$, $c_{t}^{\prime}\left(  \omega\right)  =c_{t}\left(  \omega\right)  $ for
every $\omega$ in $G_{t}\subset\Omega$, where $P\left(  G_{t}\right)  =1$ for
\emph{all} $P$ in $\mathcal{P}$.\footnote{The following perspective may be
helpful for nonspecialists in continuous time analysis. In the classical case
of a probability space $\left(  \Omega,\mathcal{F},P\right)  $ with filtration
$\{\mathcal{F}_{t}\}$, if two processes $X$ and $Y$ satisfy \textquotedblleft
for each $t$, $X_{t}=Y_{t}~$ $P$-$a.s.$\textquotedblright, then $Y$ is called
a \emph{modification} of $X $. If $P\left(  \{\omega:X_{t}=Y_{t}\text{
}\forall t\in\lbrack0,T]\}\right)  =1$, then $X$ and $Y$ are said to be
\emph{indistinguishable}. These notions are equivalent when restricted to $X$
and $Y$ having a.s. right continuous sample paths, but the second is stronger
in general. We point out, however, that the sense in which one constructs a
Brownian motion exhibiting continuous sample paths is that a suitable
modification exists (see the Kolmogorov-Centsov Theorem). A reference for the
preceding is Karatzas and Shreve \cite[pp. 2, 53]{KS}.} We abbreviate the
preceding in the form: for every $t$,
\[
c_{t}^{\prime}=c_{t}\text{ ~}\mathcal{P}\text{-}a.s.
\]
Loosely put, the latter means that the two processes are certain to yield
identical consumption levels regardless of which prior in $\mathcal{P}$ is the
true law describing the driving process. Put another way, a consumption
process as defined formally herein is portrayed in greater detail than if it
were seen from the perspective of any single prior in $\mathcal{P}$.\ For
example, if the two priors $P_{1}$ and $P_{2}$ are singular, then each
provides a description of consumption on a subset $\Omega_{i}$, $i=1,2$, of
the set of possible trajectories of the driving process, where $\Omega_{1}$
and $\Omega_{2}$ are disjoint, while using the entire set $\mathcal{P}$ yields
a description of consumption on $\Omega_{1}\cup\Omega_{2}$ and more.

\begin{remark}
If every two priors are singular, then the statement $c_{t}^{\prime}=c_{t}$
$~\mathcal{P}$-$a.s.$ amounts to standard probability $1$ statements on
disjoint parts of the state space, and thus is not far removed from a standard
equation in random variables. However, singularity is not representative - it
can be shown that the set $\mathcal{P}$ typically contains priors that, though
not equivalent, are also not mutually singular. Thus the reader is urged not
to be overly influenced by the example of singular priors which we use often
only because it starkly illustrates nonequivalence.
\end{remark}

Equations involving processes other than consumption processes are given
similar meanings. For example, the equality (\ref{Vt}) should be understood to
hold $\mathcal{P}$-almost surely for each $t$. Similar meaning is given also
to inequalities.

Utility has a range of natural properties. Most noteworthy is that the process
$\left(  V_{t}\right)  $ satisfies the recursive relation%
\begin{equation}
V_{t}=-\,\widehat{E}\left[  -\int_{t}^{\tau}\,f(c_{s},V_{s})\,ds\,-\,V_{\tau
}\,\mid\,\mathcal{F}_{t}\right]  \text{, \ }\,\,0\leq t<\tau\leq
T.\label{Vrecursive}%
\end{equation}
Such recursivity is typically thought to imply dynamic consistency. However,
the nonequivalence of priors complicates matters as we describe next.

The noted recursivity does imply the following weak form of dynamic
consistency: For any $0<\tau<T$, and any two consumption processes $c^{\prime
}$ and $c$ that coincide on $[0,\tau]$,
\[
\lbrack V_{\tau}\left(  c^{\prime}\right)  \geq V_{\tau}\left(  c\right)
\text{ \ }\mathcal{P}\text{-}a.s.]\Longrightarrow V_{0}\left(  c^{\prime
}\right)  \geq V_{0}\left(  c\right)  \text{.}%
\]
Typically, (see Duffie and Epstein \cite[p. 373]{DE} for example), dynamic
consistency is defined so as to deal also with strict rankings, that is, if
also $V_{\tau}\left(  c^{\prime}\right)  >V_{\tau}\left(  c\right)  $ on a
\textquotedblleft non-negligible\textquotedblright\ set of states, then
$V_{0}\left(  c^{\prime}\right)  >V_{0}\left(  c\right)  $. This added
requirement rules out the possibility that $c^{\prime}$ is chosen ex ante
though it is indifferent to $c$, and yet it is not implemented fully because
the individual switches to the conditionally strictly preferable $c$ for some
states at time $\tau$. The issue is how to specify \textquotedblleft
non-negligible\textquotedblright. When all priors are equivalent, then
positive probability according to any single prior is the natural
specification. However, in the absence of equivalence a similarly natural
specification is unclear. A simple illustration of the consequence is given in
the next example.

\begin{example}
[Weak Dynamic Consistency]\noindent\label{example-DC}Take $d=1$. Let the
endowment process $e$ satisfy (under $P_{0}$)%
\[
d\log e_{t}=\sigma_{t}dB_{t}\text{,}%
\]
or equivalently,
\[
de_{t}/e_{t}=\tfrac{1}{2}\sigma_{t}^{2}dt+\sigma_{t}dB_{t}\text{ \ }%
P_{0}\text{-}a.s.
\]
Here volatility is restricted only by $0<~$\underline{$\sigma$}$~\leq
\sigma_{t}\leq\overline{\sigma}$. Utility is defined, for any consumption
process $c$, by
\[
V_{0}\left(  c\right)  =\inf_{P\in\mathcal{P}}E^{P}\left[  \int_{0}%
^{T}e^{-\beta t}u\left(  c_{t}\right)  dt\right]  =\inf_{P\in\mathcal{P}%
}\left[  \int_{0}^{T}e^{-\beta t}E^{P}u\left(  c_{t}\right)  dt\right]
\text{,}%
\]
where
\[
u\left(  c_{t}\right)  =(c_{t})^{\alpha}/\alpha\text{, ~}\alpha<0\text{.}%
\]
Denote by $P^{\ast}$ the prior in $\mathcal{P}$ corresponding to $\sigma
_{t}=\overline{\sigma}$ for all $t$; that is, $P^{\ast}$ is the measure on
$\Omega$ induced by $X^{\ast}$,
\[
X_{t}^{\ast}=\overline{\sigma}B_{t}\text{, for all }t\text{ and }%
\omega\text{.}%
\]
Then
\[
V_{0}\left(  e\right)  =E^{P^{\ast}}\left[  \int_{0}^{T}e^{-\beta s}u\left(
e_{t}\right)  dt\right]  \text{.}%
\]
(This is because $u\left(  e_{t}\right)  =\alpha^{-1}\exp\left(  \alpha\log
e_{t}\right)  $ and because $\alpha<0$ makes $x\longmapsto e^{\alpha x}%
/\alpha$ concave, so that the argument in Example \ref{example-closedform} can
be adapted.)

Define the nonnegative continuous function $\varphi$\ on $\mathbb{R}$ by%
\[
\varphi(x)=\left\{
\begin{array}
[c]{cc}%
1 & x\leq\underline{\sigma}^{2}\\
\frac{2x}{\underline{\sigma}^{2}-\overline{\sigma}^{2}}-\frac
{\underline{\sigma}^{2}+\overline{\sigma}^{2}}{\underline{\sigma}%
^{2}-\overline{\sigma}^{2}}\; & \underline{\sigma}^{2}<x<\tfrac
{\underline{\sigma}^{2}+\overline{\sigma}^{2}}{2}\\
0 & \tfrac{\underline{\sigma}^{2}+\overline{\sigma}^{2}}{2}\leq x
\end{array}
\right.
\]
Fix $\tau>0$. Define the event $N_{\tau}$ by $N_{\tau}=\left\{  \omega:\langle
B\rangle_{\tau}=\underline{\sigma}^{2}\tau\right\}  $, and the consumption
process $c$ by%
\[
c_{t}=\left\{
\begin{array}
[c]{cc}%
e_{t} & 0\leq t\leq\tau\\
e_{t}+\varphi(\langle B\rangle_{\tau}/\tau) & \tau\leq t\leq T
\end{array}
\right.
\]
Then $V_{\tau}\left(  c\right)  \geq V_{\tau}\left(  e\right)  $ $\mathcal{P}%
$-almost surely and a strict preference prevails on $N_{\tau}$ because
$\varphi(\underline{\sigma}^{2})=1$\ and $P^{\underline{\sigma}}(N_{\tau}%
)=1$.{\Large \ }However, $c$ is indifferent to $e$ at time $0$ because
$\varphi(\langle B\rangle_{\tau}/\tau)=\varphi(\overline{\sigma}^{2}%
)=0${\Large \ }under $P^{\ast}$.
\end{example}

In the asset pricing application below we focus on dynamic behavior (and
equilibria) where ex ante optimal plans are implemented for all time
$\mathcal{P}$-almost surely. This requires that we examine behavior from
conditional perspectives and not only ex ante. Accordingly, if the feasible
set in the above example is $\{e,c\}$, the predicted choice would be $c$.

\section{Asset Returns\label{section-asset}}

This section describes some implications of ambiguous volatility for asset
pricing theory. First, we describe what can be said about prices based on
hedging arguments, without assuming preference maximization or equilibrium. It
is well known that ambiguous volatility leads to market incompleteness (see
Avellaneda et al. \cite{ALP}, for example) and hence that perfect hedging is
generally impossible. Accordingly, hedging arguments lead only to interval
predictions of security prices which adds to the motivation for considering
preferences and equilibrium. We explore such an equilibrium approach by
employing the utility functions defined above and a representative agent
setup. The main result is a version of the C-CAPM that applies when volatility
is ambiguous. As an illustration of the added explanatory power of the model,
it can rationalize the well documented feature of option prices whereby the
Black-Scholes implied volatility exceeds the realized volatility of the
underlying security.

In received theory, there exist positive \textquotedblleft
state-prices\textquotedblright\ that characterize arbitrage-free and
equilibrium prices in the familiar way.\footnote{We do not treat arbitrage
formally. However, at an informal level we identify no-arbitrage prices with
those produced by Black-Scholes-style hedging arguments because of their
intuitive connection and because of the formal connection that is familiar in
the standard ambiguity-free model (Duffie \cite{duffie}).} In the standard
setup where null events are defined by a reference (physical or subjective)
measure, state prices are often used as densities to define a risk neutral or
martingale measure. Densities do not apply when priors disagree about what is
possible. But, surprisingly perhaps, suitable state prices can still be derived.

There is a literature on the pricing of derivative securities when volatility
is ambiguous. The problem was first studied by Lyons \cite{Lyons} and
Avellaneda et al. \cite{ALP}; recent explorations include Denis and Martini
\cite{DM}, Cont \cite{cont} and Vorbrink \cite{vorbrink}.\footnote{These
papers often refer to uncertain volatility rather than to ambiguous
volatility.} They employ hedging arguments to derive upper and lower bounds on
security prices. However, our Theorem \ref{thm-hedge} is the first to
characterize these price bounds in terms of state prices. In addition, we
study a Lucas-style endowment economy and thus take the endowment as the basic
primitive, while the cited papers take the prices of primitive securities as
given. We are not aware of any previous studies of equilibrium in continuous
time with ambiguous volatility.

Our asset market analysis is conducted under the assumption that there is
ambiguity only about volatility and that the volatility matrix $\sigma_{t}$ is
restricted by:%
\begin{equation}
\sigma_{t}\left(  \omega\right)  \in\Gamma\text{, \ for each }t\text{ and
}\omega\text{,}\label{gamma}%
\end{equation}
where $\Gamma\subset\mathbb{R}^{d\times d}$ is compact and convex and, for all
$\sigma$ in $\Gamma$, $\sigma\sigma^{\top}\geq\hat{a}$ for some positive
definite matrix $\widehat{a}$. (In the one-dimensional case, $\Gamma=\left[
\underline{\sigma},\overline{\sigma}\right]  $ with $\underline{\sigma}>0$;
Examples \ref{example-closedform} and \ref{example-DC} both use this
specification.). The trinomial example in Section \ref{section-trinomial} is
the discrete time one-dimensional counterpart. The formal model is due to Peng
\cite{P-2006}, who calls it \textit{G-Brownian motion}, which is, loosely put,
\textquotedblleft Brownian motion with ambiGuous volatility.\textquotedblright%
\ We adopt this terminology and thus refer to the coordinate process $B$ as
being a $G$-Brownian motion under $\mathcal{P}$. Importantly, much of the
machinery of stochastic calculus, including generalizations of It\^{o}'s Lemma
and It\^{o} integration, has been extended to the framework of G-Brownian
motion (see Appendix \ref{app-GBM} for brief descriptions). This machinery is
used in the proofs. However, the proof `ideas' are standard, for example, they
exploit It\^{o}'s Lemma and a martingale representation theorem. The
difficulty is only to know when and in precisely what form such tools apply.
The statements of results do not rely on this formal material and are easy to
understand if one accepts that they differ from standard theory primarily
through the use of a new (nonadditive) notion of conditional expectation and
the substitution of \textquotedblleft$\mathcal{P}$-almost
surely\textquotedblright\ for the usual almost surely qualification.

In order to state the asset pricing results precisely, we need to be more
precise about the formal domains for random variables and stochastic
processes. They differ from the usual domains (only) because of the
nonequivalence of priors. Random payoffs occurring at a single instant are
taken to be bounded continuous functions of the state or suitable limits of
such functions. Formally, define the space $\widehat{L^{2}}(\Omega)$ to be the
completion, under the norm $\parallel\xi\parallel\equiv(\hat{E}[\mid\xi
\mid^{2}])^{\frac{1}{2}}$, of the set of all bounded continuous functions on
$\Omega$. Then $\widehat{L^{2}}(\Omega)$ is a subset of the set of measurable
random variables $\xi$ for which $\sup_{P\in\mathcal{P}}E^{P}\left(  \mid
\xi\mid^{2}\right)  <\infty$.\footnote{It contains many discontinuous random
variables. For example, $\widehat{L^{2}}(\Omega)$ contains every bounded and
lower semicontinuous function on $\Omega$ (see our companion paper).} For
processes, define $M^{2,0}$ to be the class of processes $\eta$ of the form%
\[
\eta_{t}(\omega)=%
{\displaystyle\sum\limits_{i=0}^{N-1}}
\xi_{i}(\omega)1_{[t_{i},t_{i+1})}(t),
\]
where $\xi_{i}\in\widehat{L^{2}}(\Omega)$, $0\leq i\leq N-1$, and
$0=t_{0}<\cdots<t_{N}=T$. Roughly, each such $\eta$ is a step function in
random variables from $\widehat{L^{2}}(\Omega)$. For the usual technical
reasons, we consider also suitable limits of such processes. Thus the ambient
space for processes, denoted $M^{2}$, is taken to be the completion of
$M^{2,0}$ under the norm%
\[
\parallel\eta\parallel_{M^{2}}\equiv(\hat{E}[%
{\displaystyle\int\nolimits_{0}^{T}}
\mid\eta_{t}\mid^{2}dt])^{\frac{1}{2}}.
\]
If, for every $t$, $Z_{t}=0$\ $\mathcal{P}$-$a.s.$, then $Z=0$ in $M^{2}$
(because $\hat{E}[%
{\displaystyle\int\nolimits_{0}^{T}}
\mid Z_{t}\mid^{2}dt]\leq%
{\displaystyle\int\nolimits_{0}^{T}}
\hat{E}[\mid Z_{t}\mid^{2}]dt=0$), but the converse is not valid in general.
The consumption processes $\left(  c_{t}\right)  $ and utility processes
$\left(  V_{t}\left(  c\right)  \right)  $ discussed above, as well as all
processes below related to asset markets, are taken to lie in $M^{2}$.

The domain $M^{2}$ depends on the set of priors $\mathcal{P}$, and hence also
on set $\Gamma$ from (\ref{gamma}) that describes volatility ambiguity. If
ambiguity about volatility increases in the sense that $\Gamma$ is replaced by
$\Gamma_{\ast}$, $\Gamma\subset\Gamma_{\ast}$, then it is easy to see that the
corresponding domain of processes shrinks, that is,
\begin{equation}
\Gamma\subset\Gamma_{\ast}\text{~}\Longrightarrow M_{\ast}^{2}\subset
M^{2}\text{.}\label{gamma*}%
\end{equation}
The reason is that when ambiguity increases, processes are required to be well
behaved (square integrable, for example) with respect to more probability laws.

\subsection{Hedging and state prices\label{section-sequential}}

Consider the following market environment. There is a single consumption good,
a riskless asset with return $r_{t}$ and $d$ risky securities available in
zero net supply. Returns $R_{t}$ to the risky securities are given by%
\begin{equation}
dR_{t}=b_{t}dt+s_{t}dB_{t}\text{,}\label{Rt}%
\end{equation}
where $s_{t}$ is a $d\times d$ invertible volatility matrix. Both $(b_{t})$
and $\left(  s_{t}\right)  $ are known by the investor.\footnote{$b_{t}$ and
$s_{t}$ are functions on $\Omega$, the set of possible trajectories for the
driving process. It is these \emph{functions} that are known. The trajectory
is, of course, uncertain and known only at $T$.} Define $\eta_{t}=s_{t}%
^{-1}(b_{t}-r_{t}1)$, the market price of uncertainty (a more appropriate term
here than market price of risk). It is assumed henceforth that $\left(
r_{t}\right)  $ is a bounded process in $M^{2}$; a restriction on the market
price of risk will be given below.

It is important to understand the significance of the assumption that the
returns equation holds $\mathcal{P}$-almost surely, that is, $P$-$a.s.$ for
every prior in $\mathcal{P}$. Because we are excluding ambiguity about drift,
each prior $P$ corresponds to an admissible hypothesis $\left(  \sigma
_{t}\right)  $ for volatility via (\ref{Xtheta}) and (\ref{PthetaP0}). Thus
write $P=P^{\left(  \sigma_{t}\right)  }$. Then, taking $d=1$ for simplicity,%
\[
\langle R\rangle_{t}=\int_{0}^{t}s_{\tau}^{2}d\langle B\rangle_{\tau}=\int%
_{0}^{t}s_{\tau}^{2}\sigma_{\tau}^{2}d\tau\text{ \ }P^{\left(  \sigma
_{t}\right)  }\text{-}a.s.
\]
In general, the prior implied by an alternative hypothesis $\left(  \sigma
_{t}^{\prime}\right)  $ is not equivalent to $P^{\left(  \sigma_{t}\right)  }%
$, which means that $P^{\left(  \sigma_{t}^{\prime}\right)  }$ and $P^{\left(
\sigma_{t}\right)  }$ yield different views of the quadratic variation of
returns. Consequently, the volatility of returns is ambiguous: it is certain
only that $\langle R\rangle_{t}$ lies in the interval $\left[
\underline{\sigma}^{2}\int_{0}^{t}s_{\tau}^{2}d\tau,\overline{\sigma}^{2}%
\int_{0}^{t}s_{\tau}^{2}d\tau\right]  $.

Similarly, unless explicitly stated otherwise, the $\mathcal{P}$-almost sure
qualification should be understood to apply to all other equations (and
inequalities) below even where not stated; and its significance can be
understood along the same lines.

Fix the dividend stream denoted $(\delta,\delta_{T})$, where $\delta_{t}$ is
the dividend for $0\leq t<T$ and $\delta_{T}$ is the lumpy dividend paid at
the terminal time; formally, $(\delta,\delta_{T})\in M^{2}\times
\widehat{L^{2}}(\Omega)$ . For a given time $\tau$, consider the following law
of motion for wealth on $[\tau,T]$:
\begin{align}
dY_{t}  & =(r_{t}Y_{t}+\eta_{t}^{\top}\phi_{t}-\delta_{t})dt+\phi_{t}^{\top
}dB_{t}\text{,}\;\label{superbudget}\\
Y_{\tau}  & =y\text{,}\nonumber
\end{align}
where $y$ is initial wealth, $\phi_{t}=Y_{t}s_{t}^{\top}\psi_{t}$, and
$(\psi_{t})$ is the trading strategy, that is, $\psi_{ti}$ is the proportion
of wealth invested in risky security $i$. (By nonsingularity of $s_{t}$,
choice of a trading strategy can be expressed equivalently in terms of choice
of $\left(  \phi_{t}\right)  $, which reformulation is simplifying.) Denote
the unique solution by $Y^{y,\phi,\tau}$. Define the superhedging set
\[
\mathcal{U}_{\tau}=\{y\geq0\mid\exists\phi\in M^{2}\text{ s.t. }Y_{T}%
^{y,\phi,\tau}\geq\delta_{T}\}\text{,}%
\]
and the \emph{superhedging price} $\overline{S}_{\tau}=\inf\{y\mid
y\in\mathcal{U}_{\tau}\}$. Similarly define the subhedging set%
\[
\mathcal{L}_{\tau}=\{y\geq0\mid\exists\phi\in M^{2}\text{ s.t. }Y_{T}%
^{-y,\phi,\tau}\geq-\delta_{T}\}
\]
and the \emph{subhedging price} $\underline{S}_{\tau}=\sup\{y\mid
y\in\mathcal{L}_{\tau}\}$.

The relevance of ambiguity about volatility is apparent once one realizes that
the law of motion (\ref{superbudget}), and also the inequalities at $T$ that
define $\mathcal{U}_{\tau}$ and $\mathcal{L}_{\tau}$, should be understood to
hold $\mathcal{P}$-almost surely. Thus, for example, a superhedging trading
strategy must deliver $\delta_{t}$ on $[0,T)$ and also at least $\delta_{T}$
at $T$ for all realizations that are conceiveable according to some prior.
Speaking loosely, the need to satisfy many nonequivalent priors in this way
makes superhedging difficult ($\mathcal{U}_{\tau}$ small) and the superhedging
price large. Similarly, ambiguity reduces the subhedging price. Hence the
price interval is made larger by ambiguous volatility. More precisely, by the
preceding argument, (\ref{gamma*}) and using the obvious notation,
\[
\Gamma\subset\Gamma_{\ast}\text{~}\Longrightarrow\left[  \underline{S}%
_{0},\overline{S}_{0}\right]  \subset\left[  \underline{S}_{\ast0}%
,\overline{S}_{\ast0}\right]  \text{. }%
\]
Thus an increase in volatility weakens the implications for price of a hedging
argument and naturally bolsters the case for pursuing an equilibrium analysis,
which we do after characterizing super and subhedging prices.

We show that both of the above prices can be characterized using appropriately
defined state prices. Let%
\begin{equation}
v_{t}=\underset{\varepsilon\downarrow0}{\overline{\lim}}\frac{1}{\varepsilon
}(\langle B\rangle_{t}-\langle B\rangle_{t-\varepsilon})\text{,}\label{v}%
\end{equation}
where $\overline{\lim}$ is taken componentwise.\footnote{In this we are
following Soner et. al. \cite{STZ-2010-3}. The quadratic variation process
$\langle B\rangle$ is defined in (\ref{Blim}); $v_{t}(\omega)$ takes values in
$\mathbb{S}_{d}^{>0}$, the space of all $d\times d$ positive-definite
matrices.} Under $P_{0}$, $B$ is a Brownian motion and $v_{t}$ equals the
$d\times d$ identity matrix $P_{0}$-$a.s.$ Importantly, we can also describe
$v_{t}$ as seen through the lense of any other prior in $\mathcal{P}$: if
$P=P^{\left(  \sigma_{t}\right)  }$ is a prior in $\mathcal{P}$ corresponding
via the SDE (\ref{Xtheta}) to $\left(  \sigma_{t}\right)  $,
then\footnote{Here is the proof: By Soner et al. \cite[p. 4]{STZ-2010-3},
$\langle B\rangle$ equals the quadratic variation of $B$ $P^{\left(
\sigma_{t}\right)  }$-$a.s.$; and by Oksendal \cite[p. 56]{oksendal}, the
quadratic variation of $%
{\textstyle\int\limits_{0}^{t}}
\sigma_{s}dB_{s}$ equals $%
{\textstyle\int\limits_{0}^{t}}
\sigma_{s}\sigma_{s}^{\top}ds$ \ \ $P^{\left(  \sigma_{t}\right)  }$-$a.s.$
Thus we have the $P^{\left(  \sigma_{t}\right)  }$-$a.s. $ equality in
processes $\langle B\rangle=\left(
{\textstyle\int\limits_{0}^{t}}
\sigma_{s}\sigma_{s}^{\top}ds\right)  $.$\ $ Because $\langle B\rangle$ is
absolutely continuous, its time derivative exists $a.s.$ on $[0,T]$; indeed,
the derivative at $t$ is $v_{t}$. Evidently, $\frac{d}{dt}%
{\textstyle\int\limits_{0}^{t}}
\sigma_{s}\sigma_{s}^{\top}ds=\sigma_{t}\sigma_{t}^{\top}$ for almost every
$t$. Equation (\ref{vsigma}) follows.}%
\begin{equation}
v_{t}=\sigma_{t}\sigma_{t}^{\top}\text{ ~}dt\times P^{\left(  \sigma
_{t}\right)  }\text{-}a.s.\label{vsigma}%
\end{equation}
It is assumed henceforth that $(v_{t}^{-1}\eta_{t})$ is a bounded process in
$M^{2}$.\footnote{This restriction will be confirmed below whenever $\eta$ is
taken to be endogenous.}

By the \emph{state price process} we mean the unique solution $\pi=\left(
\pi_{t}\right)  $ to
\begin{equation}
d\pi_{t}/\pi_{t}=-r_{t}dt-\eta_{t}^{\top}v_{t}^{-1}dB_{t}\text{,\ \ }\pi
_{0}=1\text{,}\label{pi}%
\end{equation}
which admits a closed form expression paralleling the classical
case:\footnote{Apply Peng \cite[Ch. 5, Remark 1.3]{P-2010}.}%
\begin{equation}
\pi_{t}=\exp\{-\int\nolimits_{0}^{t}r_{s}ds-\int\nolimits_{0}^{t}\eta
_{s}^{\top}v_{s}^{-1}dB_{s}-\tfrac{1}{2}\int\nolimits_{0}^{t}\eta_{s}^{\top
}v_{s}^{-1}\eta_{s}ds\}\text{, ~}\;0\leq t\leq T\text{.}\label{pi-solution}%
\end{equation}
We emphasize the important fact that $\pi$ is `universal' in the sense of
being defined almost surely for \emph{every} prior in $\mathcal{P}$. More
explicitly, $\pi$ satisfies: for every $t$,
\[
\pi_{t}=\exp\{-\int\nolimits_{0}^{t}r_{s}ds-\int\nolimits_{0}^{t}\eta
_{s}^{\top}\left(  \sigma_{s}\sigma_{s}^{\top}\right)  ^{-1}dB_{s}-\tfrac
{1}{2}\int\nolimits_{0}^{t}\eta_{s}^{\top}\left(  \sigma_{s}\sigma_{s}^{\top
}\right)  ^{-1}\eta_{s}ds\}\text{, \ }P^{\left(  \sigma_{t}\right)  }%
\text{-}a.s.
\]
Roughly speaking, this defines $\pi_{t}\left(  \omega\right)  $ for every
trajectory $\omega$ of the driving process that is possible according to at
least one prior in $\mathcal{P}$.\footnote{Roughly speaking, $\pi_{t}$ is
defined on the union of the supports of all priors in $\mathcal{P}$. Because
these supports need not be pairwise disjoint, it is not obvious that such a
`universal' definition exists. But (\ref{v}) and (\ref{vsigma}) ensure that
$\pi_{t}$ is well defined even where supports overlap.}

Our characterization of superhedging and subhedging prices requires an
additional arguably minor restriction on the security market. To express it,
for any $\varepsilon>0$, define
\[
\widehat{L^{2+\varepsilon}}(\Omega)=\left\{  \xi\in\widehat{L^{2}}%
(\Omega):\hat{E}[\mid\xi\mid^{2+\varepsilon}]<\infty\right\}  \text{.}%
\]
The restriction is that $\pi$ and $\delta$ satisfy {\Large \ }%
\begin{equation}
(\pi_{T}\delta_{T}+%
{\displaystyle\int\nolimits_{0}^{T}}
\pi_{t}\delta_{t}dt)\in\widehat{L^{2+\varepsilon}}(\Omega)\text{.}%
\label{pidelta}%
\end{equation}

\begin{theorem}
[Hedging prices]\label{thm-hedge}Fix a dividend stream $(\delta,\delta_{T})\in
M^{2}\times\widehat{L^{2}}(\Omega)$. Suppose that $r$ and $(v_{t}^{-1}\eta
_{t})$ are bounded processes in $M^{2}$ and that (\ref{pidelta}) is satisfied.
Then the superhedging and subhedging prices at any time $\tau$ are given by
($\mathcal{P}$-$a.s.$)%
\[
\overline{S}_{\tau}=\hat{E}[%
{\displaystyle\int\nolimits_{\tau}^{T}}
\frac{\pi_{t}}{\pi_{\tau}}\delta_{t}dt+\frac{\pi_{T}}{\pi_{\tau}}\delta
_{T}\mid\mathcal{F}_{\tau}]
\]
and
\[
\underline{S}_{\tau}=-\hat{E}[-%
{\displaystyle\int\nolimits_{\tau}^{T}}
\frac{\pi_{t}}{\pi_{\tau}}\delta_{t}dt-\frac{\pi_{T}}{\pi_{\tau}}\delta
_{T}\mid\mathcal{F}_{\tau}]\text{.}%
\]

\end{theorem}

In the special case where the security $\delta$ can be perfectly hedged, that
is, there exist $y$ and $\phi$ such that $Y_{T}^{y,\phi,0}=\delta_{T} $, then
\[
\overline{S}_{0}=\underline{S}_{0}=\hat{E}[\pi_{T}\delta_{T}+%
{\displaystyle\int\nolimits_{0}^{T}}
\pi_{t}\delta_{t}dt]=-\hat{E}[-\pi_{T}\delta_{T}-%
{\displaystyle\int\nolimits_{0}^{T}}
\pi_{t}\delta_{t}dt]\text{.}%
\]
Because this asserts equality of the supremum and infimum of expected values
of $\pi_{T}\delta_{T}+%
{\displaystyle\int\nolimits_{0}^{T}}
\pi_{t}\delta_{t}dt$ as the measures $P$ vary over $\mathcal{P}$, it follows
that\newline$E^{P}[\pi_{T}\delta_{T}+%
{\displaystyle\int\nolimits_{0}^{T}}
\pi_{t}\delta_{t}dt]$ is constant for all such measures $P$, that is, the
hedging price is unambiguous. In the further specialization where there is no
ambiguity and $P_{0}$ is the single prior, one obtains pricing by an
equivalent martingale measure whose density (on $\mathcal{F}_{t}$) with
respect to $P_{0}$ is $\pi_{t}$.

\begin{remark}
Vorbrink \cite{vorbrink} obtains an analogous characterization of hedging
prices under the assumption of $G$-Brownian motion. However, in place of our
assumption (\ref{pidelta}), he adopts the strong assumption that $b_{t}=r_{t}
$, so that the market price of uncertainty $\eta_{t}$ vanishes and $\pi
_{t}=\exp\{-\int\nolimits_{0}^{t}r_{s}ds\}$.
\end{remark}

\begin{example}
[Closed form hedging prices]\label{example-hedging}We derive the super and
subhedging prices of a European call option in a special case and compare the
results with the standard Black-Scholes formula.

Let there be one risky security ($d=1$) with price $\left(  S_{t}\right)  $
satisfying
\[
dS_{t}/S_{t}=dR_{t}=b_{t}dt+s_{t}dB_{t}.
\]
Suppose further that $r_{t}\equiv r$, $s_{t}\equiv1$ and $b_{t}-r=bv_{t}$,
where $b>0$ and $r$ are constants. Thus the market price of uncertainty is
given by $\eta_{t}=bv_{t}$, that is, using (\ref{vsigma}),
\[
\eta_{t}=b\sigma_{t}^{2}\text{ ~}dt\times P^{\left(  \sigma_{t}\right)
}\text{-}a.s.
\]
It follows that state prices are given, $\mathcal{P}$-almost surely, by%
\[
\pi_{t}=\exp\{-rt-bB_{t}-\tfrac{1}{2}b^{2}\langle B\rangle_{t}\}\text{,
~}\;0\leq t\leq T\text{.}%
\]

Consider a European call option on the risky security that matures at date $T
$ and has exercise price $K$. The super and subhedging prices at $t$ can be
written in the form $\overline{c}(S_{t},t)$ and $\underline{c}(S_{t},t)$
respectively. At the maturity date,
\[
\overline{c}(S_{T},T)=\underline{c}(S_{T},T)=\max[0,S_{T}-K]\equiv\Phi(S_{T}).
\]

By Theorem \ref{thm-hedge},
\[
\overline{c}(S_{t},t)=\hat{E}[\frac{\pi_{T}}{\pi_{t}}\Phi(S_{T})\mid
\mathcal{F}_{t}]
\]
and
\[
\underline{c}(S_{t},t)=-\hat{E}[-\frac{\pi_{T}}{\pi_{t}}\Phi(S_{T}%
)\mid\mathcal{F}_{t}]\text{.}%
\]
By the nonlinear Feynman-Kac formula in Peng \cite{P-2010}, we obtain the
following Black-Scholes-Barenblatt equation:\footnote{They reduce to the
standard Black-Scholes equation if \underline{$\sigma$}$=\overline{\sigma}$.}
\[
\partial_{t}\overline{c}+\sup_{\underline{\sigma}\leq\sigma\leq\overline
{\sigma}}\{\tfrac{1}{2}\sigma^{2}S^{2}\partial_{SS}\overline{c}\}+rS\partial
_{S}\overline{c}-r\overline{c}=0,\;\overline{c}(S,T)=\Phi(S)
\]
and%
\[
\partial_{t}\underline{c}-\sup_{\underline{\sigma}\leq\sigma\leq
\overline{\sigma}}\{-\tfrac{1}{2}\sigma^{2}S^{2}\partial_{SS}\underline{c}%
\}+rS\partial_{S}\underline{c}-r\underline{c}=0,\;\underline{c}(S,T)=\Phi(S).
\]
Because $\Phi(\cdot)$ is convex, so is $\overline{c}(\cdot,t)$.\footnote{The
argument is analogous to that in the classical Black-Scholes analysis.} It
follows that the respective suprema in the above equations are achieved at
$\overline{\sigma}$ and \underline{$\sigma$}, and we obtain%
\[
\partial_{t}\overline{c}+\tfrac{1}{2}\overline{\sigma}^{2}S^{2}\partial
_{SS}\overline{c}+rS\partial_{S}\overline{c}-r\overline{c}=0,\;\overline
{c}(S,T)=\Phi(S)
\]
and%
\[
\partial_{t}\underline{c}+\tfrac{1}{2}\underline{\sigma}^{2}S^{2}\partial
_{SS}\underline{c}+rS\partial_{S}\underline{c}-r\underline{c}%
=0,\;\underline{c}(S,T)=\Phi(S).
\]
Therefore,
\[
\overline{c}(S,t)=E^{P^{\overline{\sigma}}}[\frac{\pi_{T}}{\pi_{t}}\Phi
(S_{T}))\mid\mathcal{F}_{t}]
\]
and
\[
\underline{c}(S,t)=E^{P^{\underline{\sigma}}}[\frac{\pi_{T}}{\pi_{t}}%
\Phi(S_{T})\mid\mathcal{F}_{t}]\text{.}%
\]
In other words, the super and subhedging prices are the Black-Scholes prices
with volatilities $\overline{\sigma}$ and $\underline{\sigma}$
respectively.{\LARGE \ }
\end{example}

It is noteworthy that in contrast to this effect of volatility ambiguity, the
arbitrage-free price of a European call option is unaffected by ambiguity
about drift. This might be expected because the Black-Scholes price does not
depend on the drift of the underlying. Nevertheless some supporting detail may
be useful.

Let the security price be given as above by%
\[
dS_{t}/S_{t}=dR_{t}=(b+r)dt+dB_{t}\text{, \ }P_{0}\text{-}a.s.
\]
Model drift ambiguity by a set of priors as described in Section
\ref{section-recursive}. Each alternative hypothesis $\theta=\left(  (\mu
_{t}),1\right)  $ about the drift generates a prior $P^{\theta}$ constructed
as in (\ref{Xtheta}) and (\ref{PthetaP0}). By the Girsanov Theorem,
$X_{t}^{\theta}=\int_{0}^{\tau}(\mu_{t}dt+dB_{t})$ is a standard Brownian
motion under probability $P^{\theta}$. Therefore,
\[
dS_{t}/S_{t}=(b+r)dt+dX_{t}^{\theta}\text{, ~}P^{\theta}\text{-}a.s.
\]
and the security price follows the identical geometric Brownian motion under
$P^{\theta}$. Similarly, the counterpart of the wealth accumulation equation
(\ref{superbudget}) gives
\[
dY_{t}=(rY_{t}+\eta\phi_{t})dt+\phi_{t}dX_{t}^{\theta}\text{, ~}P^{\theta
}\text{-}a.s.\;
\]
where $\eta=b-r$. Together, the latter two equations imply that the identical
Black-Scholes price would prevail for the option regardless if $P_{0}$ or
$P^{\theta}$ were the true probability law. Because $\theta$ is an arbitrary
hypothesis about drift, it can be shown that the option price is unaffected by
ambiguity about drift.

In fact, the irrelevance of ambiguity about drift is valid much more
generally. Let there be $d$ risky securities whose prices $S_{t}\in
\mathbb{R}^{d}$ solve a stochastic differential equation of the form%
\[
dS_{t}=\widehat{b}_{t}\left(  S_{t}\right)  dt+\widehat{s}_{t}\left(
S_{t}\right)  dB_{t}\text{,}%
\]
where $\widehat{b}_{t}$ and $\widehat{s}_{t}$ are given $\mathbb{R}^{d}%
$-valued suitably well-behaved functions (for example, each $\widehat{s}%
_{t}\left(  \cdot\right)  $ is everywhere invertible). Let the instantaneous
return to the riskless security be $r_{t}\left(  S_{t}\right)  $. Finally, let
the continuous function $\psi:\mathbb{R}^{d}\rightarrow\mathbb{R}$ determine
the payoff $\psi\left(  S_{T}\right)  $ at time $T$ of a derivative security.
Under $P_{0}$, when $B$ is standard Brownian motion, the arbitrage-free price
of the derivative is defined by the Black-Scholes PDE (see Duffie \cite[Ch.
5]{duffie}, for example). The fact that the drift $\widehat{b}_{t}$ does not
enter into the PDE suggests that ambiguity about drift does not affect the
price of the derivative. Further intuition follows as above from the Girsanov
Theorem.\footnote{For any alternative hypothesis $\theta=\left(  (\mu
_{t}),1\right)  $ and corresponding prior $P^{\theta}$, the security price
process satisfies%
\[
dS_{t}=\widehat{b}_{t}\left(  S_{t}\right)  dt+\widehat{s}_{t}\left(
S_{t}\right)  dX_{t}^{\theta}\text{, }P^{\theta}\text{-}a.s.\text{,}%
\]
where $X_{t}^{\theta}=\int_{0}^{\tau}(\mu_{t}dt+dB_{t})$ is a standard
Brownian motion under $P^{\theta}$. Therefore, all conceivable truths
$P^{\theta}$ imply the identical price for the derivative, and ambiguity about
drift has no effect. Further, the corresponding hedging strategy is also
unaffected. A rigorous proof is readily constructed. We do not provide it
because ambiguity about drift alone is not our focus.} This argument covers
all the usual European options.\footnote{It is not difficult to show by a
similar argument that the arbitrage-free price of Asian options is also
unaffected by ambiguity about drift.} In contrast, as illustrated by the
example of a European call option, ambiguity about volatility does matter (the
price interval in Theorem \ref{thm-hedge} is typically nondegenerate). This is
not surprising given the known importance of volatility in option pricing.
More formally, the difference from the case of drift arises because for an
alternative hypothesis $\theta=(0,\left(  \sigma_{t}\right)  )$ for
volatility, $\left(  X_{t}^{\theta}\right)  $ satisfying $dX_{t}^{\theta
}=\sigma_{t}dB_{t}$ is \emph{not} a standard Brownian motion under $P^{\theta
}$.

\subsection{Equilibrium\label{section-eq}}

Here we use state prices to study equilibrium in a representative agent
economy with sequential security markets.

In the sequel, we limit ourselves to scalar consumption at every instant so
that $C\subset\mathbb{R}_{+}$. At the same time we generalize utility to
permit lumpy consumption at the terminal time. Thus we use the utility
functions $V_{t}$ given by%
\[
V_{t}\left(  c,\xi\right)  =-\hat{E}[-\int_{t}^{T}f(c_{s},V_{s}\left(
c,\xi\right)  )ds-u\left(  \xi\right)  \mid\mathcal{F}_{t}]\text{, ~}0\leq
t\leq T\text{,}%
\]
where $\left(  c,\xi\right)  $ varies over a subset $D$ of $M^{2}%
\times\widehat{L^{2}}(\Omega)$. Here $c$ denotes the absolutely continuous
component of the consumption process and $\xi$ is the lump of consumption at
$T$. Our analysis of utility extends to this larger domain in a
straightforward way.

When considering $\left(  c,\xi\right)  $, it is without loss of generality to
restrict attention to versions of $c$ for which $c_{T}=\xi$. With this
normalization, we can abbreviate $\left(  c,\xi\right)  =\left(
c,c_{T}\right)  $ by $c$ and identify $c$ with an element of $M^{2}%
\times\widehat{L^{2}}(\Omega)$. Accordingly write $V_{t}\left(  c,\xi\right)
$ more simply as $V_{t}\left(  c\right)  $, where\footnote{We assume the
following conditions for $f$ and $u$. (1) $f$ and $u$ are continuously
differentiable and concave. (2) There exists $\kappa>0$ such that$\mid
u_{c}(c)\mid<\kappa\left(  1+c\right)  $ ~for all $c\in C$, and $\sup\left\{
\mid f_{c}(c,V)\mid,\mid f(c,0)\mid\right\}  <\kappa\left(  1+c\right)  $ ~for
all $(c,V)\in C\times\mathbb{R}$. A consequence is that if $c\in M^{2}%
\times\widehat{L^{2}}(\Omega)$, then $u\left(  c_{T}\right)  ,~u_{c}\left(
c_{T}\right)  \in\widehat{L^{2}}(\Omega)$ and $f\left(  c_{t},0\right)
,~f_{c}\left(  c_{t},V_{t}\left(  c\right)  \right)  \in M^{2}$.}%
\begin{equation}
V_{t}\left(  c\right)  =-\hat{E}[-\int_{t}^{T}f(c_{s},V_{s}\left(  c\right)
)ds-u\left(  c_{T}\right)  \mid\mathcal{F}_{t}]\text{, ~}0\leq t\leq
T\text{.}\label{Vt-general}%
\end{equation}

The agent's endowment is given by the process $e$. Define $\pi^{e}=\left(
\pi_{t}^{e}\right)  $, called a \emph{supergradient at }$e$, by%
\begin{align}
\pi_{t}^{e}  & =\exp\left(  \int_{0}^{t}f_{v}\left(  e_{s},V_{s}\left(
e\right)  \right)  ds\right)  f_{c}\left(  e_{t},V_{t}\left(  e\right)
\right)  \text{, \ }0\leq t<T\text{,}\label{pi-e}\\
\pi_{T}^{e}  & =\exp\left(  \int_{0}^{T}f_{v}\left(  e_{s},V_{s}\left(
e\right)  \right)  ds\right)  u_{c}\left(  e_{T}\right)  \text{.}\nonumber
\end{align}

Securities, given by (\ref{Rt}) and available in zero net supply, are traded
in order to finance deviations from the endowment process $e$. Denote the
trading strategy by $(\psi_{t})$, where $\psi_{ti}$ is the proportion of
wealth invested in risky security $i$. Then wealth $Y_{t}$ evolves according
to the equation
\begin{align}
dY_{t}  & =(r_{t}Y_{t}+\eta_{t}^{\top}\phi_{t}-(c_{t}-e_{t}))dt+\phi_{t}%
^{\top}dB_{t}\text{, ~}\label{optbu2}\\
Y_{0}  & =0\text{, }c_{T}=e_{T}+Y_{T}\geq0\text{,}\nonumber
\end{align}
where $\eta_{t}=s_{t}^{-1}(b_{t}-r_{t}1)$ and $\phi_{t}=Y_{t}s_{t}^{\top}%
\psi_{t}$. We remind the reader that, by nonsingularity of $s_{t}$, choice of
a trading strategy can be expressed equivalently in terms of choice of
$\left(  \phi_{t}\right)  $.

Refer to $c$ as being \emph{feasible} if $c\in D$ and there exists $\phi$ in
$M^{2}$ such that (\ref{optbu2}) is satisfied. More generally, for any
$0\leq\tau\leq T$, consider an individual with initial wealth $Y_{\tau}$ who
trades securities and consumes during the period $[\tau,T]$. Say that $c $
\emph{is feasible on }$[\tau,T]$\emph{\ given initial wealth }$Y_{\tau} $ if
(\ref{optbu2}) is satisfied on $[\tau,T]$ and wealth at $\tau$ is $Y_{\tau}$.
Because (\ref{optbu2}) should be understood as being satisfied $P $-almost
surely for every prior in $\mathcal{P}$, greater ambiguity about volatility
tightens the feasibility restriction (paralleling the discussion in the
previous section).

State prices can be used to characterize feasible consumption plans as
described next.

\begin{theorem}
[State prices]\label{thm-pi}Define $\pi\in$ $M^{2}$ by (\ref{pi-solution}) and
let $0\leq\tau<T$.

(i) If $c$ is feasible on $[\tau,T]$ given initial wealth $Y_{\tau}$, then,
$\mathcal{P}$-$a.s.$,%
\begin{align}
Y_{\tau}  & =\hat{E}[%
{\displaystyle\int\nolimits_{\tau}^{T}}
\frac{\pi_{t}}{\pi_{\tau}}(c_{t}-e_{t})dt+\frac{\pi_{T}}{\pi_{\tau}}%
(c_{T}-e_{T})\mid\mathcal{F}_{\tau}]\label{Ytau}\\
& =-\hat{E}[-%
{\displaystyle\int\nolimits_{\tau}^{T}}
\frac{\pi_{t}}{\pi_{\tau}}(c_{t}-e_{t})dt-\frac{\pi_{T}}{\pi_{\tau}}%
(c_{T}-e_{T})\mid\mathcal{F}_{\tau}]\text{.}\nonumber
\end{align}

(ii) Conversely, suppose that (\ref{Ytau}) is satisfied and that $c_{T}\geq0
$. Then $c$ is feasible on $[\tau,T]$ given initial wealth $Y_{\tau}$.
\end{theorem}

When relevant processes are ambiguity-free diffusions, Cox and Huang
\cite{coxhuang} show that state prices can be used to transform a dynamic
process of budget constraints into a single static budget constraint. The
theorem provides a counterpart for our setting: for any given $\tau$ and
initial wealth $Y_{\tau}$, feasibility on $\left[  \tau,T\right]  $ may be
described by the `expected' expenditure constraint (\ref{Ytau}). Because both
(\ref{optbu2}) and (\ref{Ytau}) must be understood to hold $\mathcal{P}%
$-almost surely, speaking loosely, the equivalence between the dynamic and
static budget constraints is satisfied simultaneously for all hypotheses
$\left(  \sigma_{t}\right)  $ satisfying (\ref{gamma}).

Perhaps surprisingly, (\ref{Ytau}) contains two (generalized) expectations.
Their conjunction can be interpreted as in the discussion following Theorem
\ref{thm-hedge}: in expected value terms any feasible consumption plan
\emph{unambiguously} (that is, for every prior) exhausts initial wealth when
consumption is priced using $\pi$. Such an interpretation is evident when
$\tau=0$; a similar interpretation can be justified when $\tau>0$.

Turn to equilibrium. Say that $(e,\left(  r_{t},\eta_{t}\right)  )$ is a
\emph{sequential equilibrium} if for every $c$: For each $\tau$, $\mathcal{P}%
$-almost surely,
\[
c\in \Upsilon_{\tau}\left(  0\right)  \Longrightarrow V_{\tau}\left(  c\right)
\leq V_{\tau}\left(  e\right)  \text{.}%
\]
Thus equilibrium requires not only that the endowment $e$ be optimal at time
$0$, but also that it remain optimal at any later time given that $e$ has been
followed to that point.

The main result of this section follows.

\begin{theorem}
[Sequential Equilibrium I]\label{thm-eqI}Define $\pi,\pi^{e}\in$ $M^{2}$ by
(\ref{pi-solution}) and (\ref{pi-e}) respectively, and assume that
$\mathcal{P}$-almost surely,
\begin{equation}
\pi_{t}^{e}/\pi_{0}^{e}=\pi_{t}\text{.}\label{pipie}%
\end{equation}
Then $(e,\left(  r_{t},\eta_{t}\right)  )$ is a sequential equilibrium.
\end{theorem}

Condition (\ref{pipie}) is in the spirit of the Duffie and Skiadas \cite{DS}
approach to equilibrium analysis (see also Skiadas \cite{skiadas2008} for a
comprehensive overview of this approach). Speaking informally, the process
$\pi^{e}/\pi_{0}^{e}$ describes marginal rates of substitution at $e$, while
$\pi$ describes trade-offs offered by the market. Their equality relates the
riskless rate and the market price of risk to consumption and continuation
utility through the equation%
\[
d\pi_{t}^{e}/\pi_{t}^{e}=-r_{t}dt-\eta_{t}^{\top}v_{t}^{-1}dB_{t}\text{.}%
\]

To be more explicit, suppose that $e$ satisfies
\[
de_{t}/e_{t}=\mu_{t}^{e}dt+(s_{t}^{e})^{\top}dB_{t}\text{.}%
\]
Consider also two specific aggregators. For the standard aggregator
(\ref{fstandard}),
\begin{equation}
\pi_{t}^{e}=\exp\left(  -\beta t\right)  u_{c}\left(  e_{t}\right)  \text{,
\ }0\leq t\leq T\text{.}\label{pi-e2}%
\end{equation}
Then Ito's Lemma for $G$-Brownian motion (Appendix \ref{app-GBM}) and
(\ref{vsigma}) imply that%
\begin{equation}
b_{t}-r_{t}\mathbf{1}=s_{t}\eta_{t}=-\left(  \tfrac{e_{t}u_{cc}\left(
e_{t}\right)  }{u_{c}\left(  e_{t}\right)  }\right)  s_{t}\sigma_{t}\sigma
_{t}^{\top}s_{t}^{e}\text{, \ }P^{\left(  \sigma_{t}\right)  }\text{-}%
a.s.\label{returns}%
\end{equation}
which is a version of the C-CAPM for our setting.\footnote{Equality here (and
in similar equations below) means that the two processes $(b_{t}-r_{t}1)$\ and
$(-\left(  \tfrac{e_{t}u_{cc}\left(  e_{t}\right)  }{u_{c}\left(
e_{t}\right)  }\right)  s_{t}\sigma_{t}\sigma_{t}^{\top}s_{t}^{e})$\ are equal
as processes in $M^{2}$.\ Notice also that $v_{t}^{-1}\eta_{t}=-\left(
\tfrac{e_{t}u_{cc}\left(  e_{t}\right)  }{u_{c}\left(  e_{t}\right)  }\right)
s_{t}^{e}$, yielding a process in $M^{2}$ and thus confirming our prior
assumption on security markets.} From the perspective of the measure $P_{0}$
according to which the coordinate process $B$ is a Brownian motion,
$\sigma_{t}\sigma_{t}^{\top}$ is the identity matrix and one obtains the usual
C-CAPM. However, speaking informally, the equation (\ref{returns}) relates
excess returns to consumption also along trajectories that are consistent with
alternative hypotheses $\left(  \sigma_{t}\right)  $ about the nature of the
driving process.

The other case is the so-called Kreps-Porteus aggregator (Duffie and Epstein
\cite{DE}). Let
\begin{equation}
f(c,v)=\frac{c^{\rho}-\beta(\alpha v)^{\rho/\alpha}}{\rho(\alpha
v)^{(\rho-\alpha)/\alpha}}\text{,}\label{KPagg}%
\end{equation}
where $\beta\geq0$ and $0\not =\rho,\alpha\leq1$; $\left(  1-\alpha\right)  $
is the measure of relative risk aversion and $\left(  1-\rho\right)  ^{-1}$ is
the elasticity of intertemporal substitution. To evaluate terminal lumpy
consumption as in (\ref{Vt-general}), we take
\[
u\left(  c_{t}\right)  =(c_{t})^{\alpha}/\alpha\text{, ~}0\not =%
\alpha<1\text{.}%
\]
The implied version of the C-CAPM is%
\begin{equation}
b_{t}-r_{t}1=\rho^{-1}[\alpha(1-\rho)s_{t}\sigma_{t}\sigma_{t}^{\top}s_{t}%
^{e}+(\rho-\alpha)s_{t}\sigma_{t}\sigma_{t}^{\top}s^{M}],\;P^{\left(
\sigma_{t}\right)  }\text{-}a.s.\label{returnsKP}%
\end{equation}
where $s^{M}$ is the volatility of wealth in the sense that%
\[
dY_{t}/Y_{t}=b^{M}dt+(s^{M})^{\top}dB_{t}\text{, ~}\mathcal{P}\text{-}a.s.
\]
In the absence of ambiguity where $P_{0}$ alone represents beliefs, then
(\ref{returnsKP}) reduces to the two-factor model of excess returns derived by
Duffie and Epstein \cite{DE-RFS}.

Equation (\ref{returnsKP}) is derived in Appendix \ref{app-KP}, which also
presents a result for general aggregators.

\subsection{Minimizing priors\label{section-minimizing}}

Equation (\ref{pipie}) is a sufficient condition for sequential equilibrium.
Here we describe an alternative route to equilibrium that is applicable under
an added assumption and that yields an alternative form of C-CAPM.

The intuition for what follows is based on a well known consequence of the
minimax theorem for multiple priors utility in abstract environments. For
suitable optimization problems, if a prospect, say $e$, is feasible and if the
set of priors contains a worst-case scenario $P^{\ast}$ for $e$, then $e $ is
optimal if and only if it is optimal also for a Bayesian agent who uses the
single prior $P^{\ast}$. Moreover, by a form of envelope theorem, $P^{\ast}$
suffices to describe marginal rates of substitution at $e$ and hence also
supporting shadow prices. This suggests that there exist sufficient conditions
for $e$ to be part of an equilibrium in our setup that refer to $P^{\ast}$ and
less extensively to all other priors in $\mathcal{P} $. We proceed now to
explore this direction.

As a first step, define $P^{\ast}\in\mathcal{P}$ to be a \emph{minimizing
measure for }$e$ if%
\begin{equation}
V_{0}\left(  e\right)  =E^{P^{\ast}}[\int_{0}^{T}f(e_{s},V_{s}\left(
e\right)  )ds+u\left(  e_{T}\right)  ]\text{.}\label{min}%
\end{equation}
As discussed when defining equilibrium, the fact that only weak dynamic
consistency is satisfied requires that one take into account also conditional
perspectives. Speaking informally, a minimizing measure $P^{\ast}$ as above
need not be minimizing conditionally at a later time because of the
nonequivalence of priors and the uncertainty about what is possible. (Example
\ref{example-DC} is readily adapted to illustrate this.) Thus to be relevant
to equilibrium, a stronger notion of \textquotedblleft
minimizing\textquotedblright\ is required.

Recall that for any prior $P$ in $\mathcal{P}$, $P_{\tau}^{\omega}$ is the
version of the regular conditional of $P$; importantly, it is well-defined for
every $\left(  \tau,\omega\right)  $. Say that $P^{\ast}\in$ $\mathcal{P}$ is
a \emph{dynamically minimizing measure for }$e$ if, for all $\tau$,
$\mathcal{P}$-$a.s.$,
\begin{equation}
V_{\tau}\left(  e\right)  =E^{(P^{\ast})_{\tau}^{\omega}}[\int_{\tau}%
^{T}f(e_{s},V_{s}\left(  e\right)  )ds+u\left(  e_{T}\right)  ]\text{.}%
\label{dynmin}%
\end{equation}

Next relax the equality (\ref{pipie}) and assume instead: For every $\tau$ and
$\mathcal{P}$-almost surely in $\omega$,
\begin{equation}
\pi_{t}^{e}/f_{c}\left(  e_{0},V_{0}\left(  e\right)  \right)  =\pi_{t}\text{
on }[\tau,T]\text{ ~}(P^{\ast})_{\tau}^{\omega}\text{-}a.s.\label{pipie2}%
\end{equation}
Note that equality is assumed not only ex ante $P^{\ast}$-$a.s.$ but also
conditionally, even conditioning on events that are $P^{\ast}$-null but that
are possible according to other priors in $\mathcal{P}$.

\begin{theorem}
[Sequential equilibrium II]\label{thm-eqII}Let $P^{\ast}$ be a dynamic
minimizer for $e$ and assume (\ref{pipie2}). Then $(e,\left(  r_{t},\eta
_{t}\right)  )$ is a sequential equilibrium.
\end{theorem}

The counterparts of the C-CAPM relations (\ref{returns}) and (\ref{returnsKP})
are:\footnote{Assume that there exists a dynamic minimizer $P^{\ast}$ for $e$.
Then one can show that (\ref{returns}) implies (\ref{returns2}) and
(\ref{returnsKP}) implies (\ref{returnsKP2}).} For every $\tau$, $\mathcal{P}%
$-$a.s.$,
\begin{equation}
b_{t}-r_{t}\mathbf{1}=-\left(  \tfrac{e_{t}u_{cc}\left(  e_{t}\right)  }%
{u_{c}\left(  e_{t}\right)  }\right)  s_{t}(\sigma_{t}^{\ast}\sigma_{t}%
^{\ast\top})s_{t}^{e}\text{ \ on }[\tau,T]\text{ ~}(P^{\ast})_{\tau}^{\omega
}\text{-}a.s.\label{returns2}%
\end{equation}
and
\begin{equation}
b_{t}-r_{t}1=\rho^{-1}[\alpha(1-\rho)s_{t}(\sigma_{t}^{\ast}\sigma_{t}%
^{\ast\top})s_{t}^{e}+(\rho-\alpha)s_{t}(\sigma_{t}^{\ast}\sigma_{t}^{\ast
\top})s^{M}\text{ \ on }[\tau,T]\text{ ~}(P^{\ast})_{\tau}^{\omega}%
\text{-}a.s.\label{returnsKP2}%
\end{equation}
Here $P^{\ast}=P^{\left(  \sigma_{t}^{\ast}\right)  }$ is induced by the
process $\left(  \sigma_{t}^{\ast}\right)  $ as in (\ref{Xtheta}).

There are several `nonstandard' features of these relations that we interpret
in the following example where the equations take on a more concrete form.
However, it may be useful to consider the general forms briefly. For
simplicity, consider (\ref{returns2}) corresponding to the standard
aggregator. One effect of ambiguous volatility is that the relevant
instantaneous covariance between asset returns and consumption is modified
from $s_{t}s_{t}^{e}$ to $s_{t}(\sigma_{t}^{\ast}\sigma_{t}^{\ast\top}%
)s_{t}^{e}$, where $\left(  \sigma_{t}^{\ast}\right)  $ is the worst-case
hypothesis for volatility. This adjustment reflects a conservative attitude
and confidence only that volatility $\left(  \sigma_{t}\right)  $ lies
everywhere in $\Gamma$ rather than in any single hypothesis, such as
$\sigma_{t}\equiv1$, satisfying this constraint.

Compare (\ref{returns2}) also with the C-CAPM relation derived assuming
ambiguity about drift only. In that case, Chen and Epstein \cite{CE} show
that, instead of (\ref{returns2}), mean excess returns satisfy%
\begin{equation}
b_{t}-r_{t}\mathbf{1}=-\left(  \tfrac{e_{t}u_{cc}\left(  e_{t}\right)  }%
{u_{c}\left(  e_{t}\right)  }\right)  s_{t}s_{t}^{e}\text{ }+s_{t}\mu
_{t}^{\ast}\text{ }\ P_{0}\text{-}a.s.\label{returns3}%
\end{equation}
where $\left(  \mu_{t}^{\ast}\right)  $ is the worst-case hypothesis for
drift.\footnote{More precisely, in the notation of Section
\ref{section-priors}, $P^{\left(  (\mu_{t}^{\ast}),1\right)  }$ is a minimizer
in the utility calculation $V_{0}\left(  e\right)  =\inf_{P\in\mathcal{P}%
}E^{P}[\int_{0}^{T}u(e_{s})e^{-\beta s}ds]$.} It is difficult to compare these
two alternative adjustments for ambiguity in general qualitative terms.
Presumably, each kind of ambiguity matters in some contexts, (though recall
that drift ambiguity has no effect in European options markets). Because both
kinds of ambiguity may matter simultaneously, one obviously would like to
establish a version of C-CAPM that accommodates both. However, that would
require extensions of the machinery described in Appendix \ref{app-GBM} that
to our knowledge is currently available only for environments described by
$G$-Brownian motion.

Further interpretation and comparisons are discussed in the context of a final example.

\subsection{A final example\label{section-finalexample}}

Theorem \ref{thm-eqI} begs the question whether or when dynamic minimizers
exist. We have no general answers at this point. But they exist in the
following example. Its simplicity also helps to illustrate the effects of
ambiguous volatility on asset returns.

We build on previous examples. Let $d\geq1$. The endowment process $e$
satisfies (under $P_{0}$)%
\begin{equation}
d\log e_{t}=(s^{e})^{\top}\sigma_{t}dB_{t}\text{, ~}e_{0}>0\text{
given,}\label{e}%
\end{equation}
where $s^{e}$ is constant and $B$ is a G-Brownian motion (thus the volatility
matrix $\sigma_{t}$ is restricted only to lie in $\Gamma$). We assume that
$P_{0}$ lies in $\mathcal{P}$, that is, $\Gamma$ admits the constant $d\times
d$ identity matrix. Utility is defined, for any consumption process $c$, by
the following special case of (\ref{Vt-general}):
\[
V_{t}\left(  c\right)  =-\hat{E}[-\int_{t}^{T}u(c_{s})e^{-\beta(s-t)}%
ds-e^{-\beta(T-t)}u\left(  c_{T}\right)  \mid\mathcal{F}_{t}]\text{,}%
\]
where the felicity function $u$ is
\[
u\left(  c_{t}\right)  =(c_{t})^{\alpha}/\alpha\text{, ~}0\not =%
\alpha<1\text{.}%
\]

There exists a dynamic minimizer for $e$ that depends on the sign of $\alpha$.
Compute that%
\[
u\left(  e_{t}\right)  =\alpha^{-1}e_{t}^{\alpha}=\alpha^{-1}e_{0}^{\alpha
}\exp\left\{  \alpha\int_{0}^{t}(s^{e})^{\top}\sigma_{s}dB_{s}\right\}
\]
Let \underline{$\sigma$}\ and $\overline{\sigma}$ solve respectively
\begin{equation}
\min_{\sigma\in\Gamma}tr\left(  \sigma\sigma^{\top}s^{e}(s^{e})^{\top}\right)
\text{ and}\ \max_{\sigma\in\Gamma}tr\left(  \sigma\sigma^{\top}s^{e}%
(s^{e})^{\top}\right)  .\label{sigmabar}%
\end{equation}
If $d=1$, then $\Gamma$ is a compact interval and \underline{$\sigma$} and
$\overline{\sigma}$ are its left and right endpoints. Let $P^{\ast}$\ be the
measure on $\Omega$\ induced by $P_{0}$ and $X^{\ast}$, where%
\[
X_{t}^{\ast}=\overline{\sigma}^{\top}B_{t}\text{, for all }t\text{ and }%
\omega\text{;}%
\]
define $P^{\ast\ast}$\ similarly using \underline{$\sigma$} and $X^{\ast\ast}%
$. Then, by a slight extension of the observation in Example
\ref{example-closedform}, $P^{\ast}$\ is a dynamic minimizer for $e$\ if
$\alpha<0$\ and $P^{\ast\ast}$\ is a dynamic minimizer for $e$\ if $\alpha>0$.

\begin{remark}
That the minimizing measure corresponds to constant volatility is a feature of
this example. More generally, the minimizing measure in $\mathcal{P}$ defines
a specific stochastic volatility model. It is interesting to note that when
volatility is modeled by robustifying the Hull-White and Heston parametric
forms, for example, the minimizing measure does not lie in either parametric
class. Rather it corresponds to pasting the two alternatives together
endogenously, that is, in a way that depends on the endowment process and on
$\alpha$.
\end{remark}

We describe further implications assuming $\alpha<0$; the corresponding
statements for $\alpha>0$ will be obvious to the reader. Interpretation of the
sign of $\alpha$ is confounded by the dual role of $\alpha$ in the additive
expected utility model. However, the example can be generalized to the
Kreps-Porteus aggregator (\ref{KPagg}) and then the same characterization of
the worst-case volatility is valid with $1-\alpha$ interpretable as the
measure of relative risk aversion. Therefore, the intuition is clear for the
pricing results that follow: only the largest (in the sense of (\ref{sigmabar}%
)) volatility $\overline{\sigma}$ is relevant assuming $\alpha<0$ because it
represents the worst-case scenario given a large (greater than $1$) measure of
relative risk aversion.

Corresponding regular conditionals have a simple form. For example, $(P^{\ast
})_{\tau}^{\omega}$ is the measure on $\Omega$ induced by the stochastic
differential equation (under $P_{0}$)%
\[
\left\{
\begin{array}
[c]{rl}%
dX_{t} & =\overline{\sigma}dB_{t},\;\tau\leq t\leq T\\
X_{t} & =\omega_{t},\;0\leq t\leq\tau
\end{array}
\right.
\]
Thus under $(P^{\ast})_{\tau}^{\omega}$, the increment $B_{t}-B_{\tau}$ is
$N\left(  0,\overline{\sigma}\overline{\sigma}^{\top}\left(  t-\tau\right)
\right)  $ for $\tau\leq t\leq T$.

The C-CAPM (\ref{returns2}) takes the form (assuming $\alpha<0$): For every
$\tau$, $\mathcal{P}$-almost surely in $\omega$,
\[
b_{t}-r_{t}\mathbf{1}=\left(  1-\alpha\right)  s_{t}(\overline{\sigma
}\overline{\sigma}^{\top})s^{e}\text{ \ on }[\tau,T]\text{ ~}(P^{\ast})_{\tau
}^{\omega}\text{-}a.s.
\]
For comparison purposes, it is convenient to express this equation partially
in terms of $P_{0}$. The measures $P_{0}$ and $P^{\ast}$ differ only via the
change of variables defined via the SDE (\ref{Xtheta}). Therefore, we arrive
at the following equilibrium condition: \textit{For every }$\tau$%
,\textit{\ }$\mathcal{P}$\textit{-almost surely in }$\omega$,
\begin{equation}
\widehat{b}_{t}-\widehat{r_{t}}\mathbf{1}=\left(  1-\alpha\right)
\widehat{s_{t}}(\overline{\sigma}\overline{\sigma}^{\top})s^{e}\text{ \ on
}[\tau,T]\text{ ~}(P_{0})_{\tau}^{\omega}\text{-}%
a.s.\label{returns-example-hat}%
\end{equation}
where $\widehat{b}_{t}=b_{t}\left(  X_{\cdot}^{\overline{\sigma}}\right)  $,
$\widehat{r}_{t}=r_{t}\left(  X_{\cdot}^{\overline{\sigma}}\right)  $ and
$\widehat{s}_{t}=s_{t}\left(  X_{\cdot}^{\overline{\sigma}}\right)  $,
corresponding to the noted change of variables under which $B_{t}\longmapsto
X_{t}^{\overline{\sigma}}=\overline{\sigma}^{\top}B_{t}$. Note that the
difference between random variables with and without hats is ultimately not
important because they follow identical distributions under $(P_{0})_{\tau
}^{\omega}$ and $(P^{\ast})_{\tau}^{\omega}$ respectively.

The impact of ambiguous volatility is most easily seen by comparing with the
standard C-CAPM obtained assuming complete confidence in the single
probability law $P_{0}$ which renders $B$ a standard Brownian motion. Then the
prediction for asset returns is%
\[
b_{t}-r_{t}\mathbf{1}=\left(  1-\alpha\right)  s_{t}s^{e}\text{ \ on
}[0,T]\text{ ~}P_{0}\text{-}a.s.
\]
or equivalently: \textit{For every }$\tau$\textit{\ and }$P_{0}$%
\textit{-almost surely in }$\omega$,%
\begin{equation}
b_{t}-r_{t}\mathbf{1}=\left(  1-\alpha\right)  s_{t}s^{e}\text{ \ on }%
[\tau,T]\text{ ~}(P_{0})_{\tau}^{\omega}\text{-}a.s.\label{returnsP0}%
\end{equation}

There are two differences between the latter and the equilibrium condition
(\ref{returns-example-hat}) for our model. First the \textquotedblleft
instantaneous covariance\textquotedblright\ between asset returns and
consumption is modified from $s_{t}s^{e}$ to $\widehat{s_{t}}(\overline
{\sigma}\overline{\sigma}^{\top})s^{e}$ reflecting the fact that
$\overline{\sigma}$ is the worst-case volatility scenario for the
representative agent. Such an effect, whereby ambiguity leads to standard
equilibrium conditions except that the reference measure is replaced by the
worst-case measure, is familiar from the literature. The second difference is
new. Condition (\ref{returnsP0}) refers to the single measure $P_{0}$ only and
events that are null under $P_{0}$ are irrelevant.\footnote{Similarly for the
C-CAPM (\ref{returns3}) when only drift is ambiguous, because then all priors
are equivalent to $P_{0}$.} In contrast, the condition
(\ref{returns-example-hat}) is required to hold $\mathcal{P}$\textit{-}almost
surely in $\omega$ because, as described in Example \ref{example-DC}, dynamic
consistency requires that possibility be judged according to all priors in
$\mathcal{P}$.

Turn to a brief consideration of corresponding equilibrium prices. Fix a
dividend stream $(\delta,\delta_{T})\in M^{2}\times\widehat{L^{2}}(\Omega)$
where the security is available in zero net supply. Then its equilibrium price
$S^{\delta}=\left(  S_{\tau}^{\delta}\right)  $ is given by: For all $\tau$,
$\mathcal{P}$\textit{-}$a.s.$ in $\omega$,
\begin{equation}
S_{\tau}^{\delta}=E^{(P^{\ast})_{\tau}^{\omega}}[%
{\displaystyle\int\nolimits_{\tau}^{T}}
\frac{\pi_{t}^{e}}{\pi_{\tau}^{e}}\delta_{t}dt+\frac{\pi_{T}^{e}}{\pi_{\tau
}^{e}}\delta_{T}]\text{,}\label{Stau}%
\end{equation}
which lies between the hedging bounds in Theorem \ref{thm-hedge} by
(\ref{pipie2}).\footnote{The proof is analogous to that of Lemma
\ref{lemma-eq}, particularly surrounding (\ref{pasting}).} It is interesting
to compare this equilibrium pricing rule with the price bounds derived from
hedging arguments (Theorem \ref{thm-hedge}). Suppose the security in question
is an option on an underlying. Under the conditions of Example
\ref{example-hedging}, the volatilities used to define the upper and lower
price bounds depend on whether the terminal payoff is (globally) convex or
concave as a function of the price of the underlying. In contrast, the
volatility used for equilibrium pricing is the same for all options (and other
securities) and depends only on the endowment and the preference parameter
$\alpha$. This difference is further illustrated below.

If $\delta=e$, then elementary calculations yield the time $\tau$ price of the
endowment stream in the form%
\[
S_{\tau}^{e}=A_{\tau}e_{\tau}=A_{\tau}e_{0}\exp\left(  (s^{e})^{\top}B_{\tau
})\right)  ,
\]
where $A_{\tau}>0$ is deterministic and $A_{T}=1$.\ Thus $\log\left(  S_{\tau
}^{e}/A_{\tau}\right)  =\log e_{\tau}$ and the logarithm of (deflated) price
is also a G-Brownian motion. We can also price an option on the endowment.
Thus let $\delta_{t}=0$ for $0\leq t<T$ and $\delta_{T}=\psi\left(  S_{T}%
^{e}\right)  $. Denote its price process by $S^{\psi}$. From (\ref{Stau}), any
such derivative is priced in equilibrium as though $\sigma_{t}$ were constant
at $\overline{\sigma}$ (or at \underline{$\sigma$} if $\alpha>0$). In
particular, for a European call option where $\delta_{T}=\left(  S_{T}%
^{e}-\kappa\right)  ^{+}$, its equilibrium price at $\tau$ is $BS_{\tau
}\left(  (s^{e})^{\top}\overline{\sigma},T,\kappa\right)  $, where the latter
term denotes the Black-Scholes price at $\tau$ for a call option with strike
price $\kappa$ and expiry time $T$ when the underlying security price process
is geometric Brownian motion with volatility $(s^{e})^{\top}\overline{\sigma}%
$. Thus the Black-Scholes implied variance is $tr\left(  \overline{\sigma
}\overline{\sigma}^{\top}s^{e}(s^{e})^{\top}\right)  $ which exceeds every
conceivable realized variance $tr\left(  \sigma\sigma^{\top}s^{e}(s^{e}%
)^{\top}\right)  $, $\sigma\in\Gamma$, consistent with a documented empirical
feature of option prices.

\section{Concluding Remarks\label{section-conclude}}

We have described a model of utility over continuous time consumption streams
that can accommodate ambiguity about volatility. Such ambiguity necessitates
dropping the assumption that a single measure defines null events, which is a
source of considerable technical difficulty. The economic motivation provided
for confronting the technical challenge is the importance of stochastic
volatility modeling in both financial economics and macroeconomics, the
evidence that the dynamics of volatility are complicated and difficult to pin
down empirically, and the presumption that complete confidence in any single
parametric specification is unwarranted and implausible. (Recall, for example,
the quote in the introduction from Carr and Lee \cite{carrlee}.) These
considerations suggest the potential usefulness of `robust stochastic
volatility' models (Section \ref{section-why}). We have shown that important
elements of representative agent asset pricing theory extend to an environment
with ambiguous volatility. We also provided one example of the added
explanatory power of ambiguous volatility - it gives a way to understand the
documented feature of option prices whereby the Black-Scholes implied
volatility exceeds the realized volatility of the underlying security.
However, a question that remains to be answered more broadly and thoroughly is
\textquotedblleft does ambiguity about volatility and possibility matter
empirically?\textquotedblright\ In particular, it remains to determine the
empirical content of the derived C-CAPM relations. The contribution of this
paper has been to provide a theoretical framework within which one could
address such questions.

There are also several extensions at the theoretical level that seem worth
pursuing. The utility formulation should be generalized to environments with
jumps, particularly in light of the importance attributed to jumps for
understanding options markets. The asset market analysis should be extended to
permit ambiguity specifications more general than $G$-Brownian motion.
Extension to heterogeneous agent economies is important and intriguing. The
nonequivalence of measures raises questions about existence of equilibrium and
about the nature of no-arbitrage pricing (for reasons discussed in Willard and
Dybvig \cite{wd}).

Two further questions that merit attention are more in the nature of
refinements, albeit nontrivial ones and beyond the scope of this paper. The
fact that utility is recursive but not strictly so suggests that though not
every time $0$ optimal plan may be pursued subsequently at all relevant nodes,
one might expect that (under suitable regularity conditions) there exists at
least one time $0$ optimal plan that will be implemented. (This is the case in
Example \ref{example-DC}\ and also in the asset market example in Section
\ref{section-finalexample}.) Sufficient conditions for such existence should
be explored. Secondly, Sections \ref{section-minimizing} and
\ref{section-finalexample} demonstrated the significance of worst-case
scenarios in the form of dynamic minimizing measures. Their existence and
characterization pose important questions.

In terms of applications, we note that the model (slightly modified) can be
interpreted in terms of investor sentiments. Replace all infima by suprema and
vice versa. Then, the consumer may be described as an ambiguity lover, or
alternatively in terms of optimism and overconfidence. For example, in a
recent study of how the pricing kernel is affected by sentiment, Barone-Adesi
et al. \cite{shefrin} subdivide the latter and define optimism as occurring
when the investor overestimates mean returns and overconfidence as occurring
when return volatility is underestimated. This fits well with the distinction
we have emphasized at a formal modeling level between ambiguity about drift
and ambiguity about volatility. In a continuous time setting, ambiguity about
drift, or optimism, can be modeled in a probability space framework, but not
so ambiguity about volatility, or overconfidence.\footnote{The applied finance
literature has not used sets of priors in modeling sentiment. The use of sets
gives a best scenario, or subjective prior, that depends on the portfolio
being evaluated. Thus optimism can be exhibited for every portfolio as one
might expect of an investor who has an optimistic nature. In contrast when the
subjective prior is fixed, then a high estimated return for a security implies
pessimism when the agent considers going short.}

We mention one more potential application. Working in a discrete-time setting,
Epstein and Schneider \cite{es2008} point to ambiguous volatility as a way to
model signals with ambiguous precision. This leads to a new way to measure
information quality that has interesting implications for financial models
(see also Illeditsch \cite{illeditsch}). The utility framework that we provide
should permit future explorations of this dimension of information quality in
continuous time settings.

\appendix

\section{Appendix: G-Brownian Motion\label{app-GBM}}

Peng \cite{P-2006} introduced $G$-Brownian motion using PDE's (specifically, a
nonlinear heat equation). Further contributions are due to Denis et al.
\cite{DHP} and Soner et al. \cite{STZ}. For the convenience of the reader, in
this appendix we outline some key elements of the theory of $G$-Brownian
motion in terms of the specifics of our model.

\medskip

\noindent\textit{It\^{o} Integral and Quadratic Variation Process}: For each
$\eta\in M^{2}$, we can consider the usual It\^{o} integral $\int%
\nolimits_{0}^{T}\eta_{t}^{\intercal}dB_{t}$, which lies in $\widehat{L^{2}%
}(\Omega)$. Each $P\in\mathcal{P}$ provides a different perspective on the
integral; a comprehensive view requires that one consider all priors. The
quadratic variation process $\langle B\rangle$ also agrees with the usual
quadratic variation process $\mathcal{P}$-$a.s.$ In Section
\ref{section-sequential} we defined a universal process $v$ (via (\ref{v}))
and proved that
\[
\langle B\rangle=\left(
{\textstyle\int\limits_{0}^{t}}
v_{s}ds:0\leq t\leq T\right)  \text{.}%
\]

The following properties are satisfied for any $\lambda,\eta\in M^{2}$,
$X\in\widehat{L^{2}}(\Omega_{T})$ and constant $\alpha$:
\[%
\begin{array}
[c]{l}%
\hat{E}[B_{t}]=0\text{, }\hat{E}[\int\nolimits_{0}^{T}\eta_{t}^{\top}%
dB_{t}]=0\text{,}\\
\hat{E}[(\int\nolimits_{0}^{T}\eta_{t}^{\top}dB_{t})^{2}]=\hat{E}%
[\int\nolimits_{0}^{T}\eta_{t}^{\top}v_{t}\eta_{t}dt]\text{,}\\
\int\nolimits_{0}^{T}(\alpha\eta_{t}^{\top}+\lambda_{t}^{\top})dB_{t}%
=\alpha\int\nolimits_{0}^{T}\eta_{t}^{\top}dB_{t}+\int\nolimits_{0}^{T}%
\lambda_{t}^{\top}dB_{t}\text{ ~}q.s.\\
\hat{E}[X+\int\nolimits_{s}^{T}\eta_{t}^{\top}dB_{t}\mid\mathcal{F}_{s}%
]=\hat{E}[X\mid\mathcal{F}_{s}]+\hat{E}[\int\nolimits_{s}^{T}\eta_{t}^{\top
}dB_{t}\mid\mathcal{F}_{s}]=\hat{E}[X\mid\mathcal{F}_{s}]
\end{array}
\]
For the one dimensional case ($\Gamma=[\underline{\sigma},\bar{\sigma}]$,
$\underline{\sigma}>0$), we have%
\[%
\begin{array}
[c]{l}%
\underline{\sigma}^{2}t\leq\hat{E}[(B_{t})^{2}]\leq\bar{\sigma}^{2}t\text{,}\\
\underline{\sigma}^{2}\hat{E}[\int\nolimits_{0}^{T}\eta_{t}^{2}dt]\leq\hat
{E}[(\int\nolimits_{0}^{T}\eta_{t}dB_{t})^{2}]\leq\bar{\sigma}^{2}\hat{E}%
[\int\nolimits_{0}^{T}\eta_{t}^{2}dt]\text{.}%
\end{array}
\]

\medskip

\noindent\textit{It\^{o}'s Formula}: Consider%

\[
X_{t}=X_{0}+\int\nolimits_{0}^{t}\alpha_{s}ds+\int\nolimits_{0}^{t}\gamma
_{s}dB_{s}%
\]
where $\alpha$ and $\gamma$ are in $M^{2}(\mathbb{R}^{d})$ and $M^{2}%
(\mathbb{R}^{d\times d})$ respectively. (Define $M^{2}(\mathbb{R}^{\ell\times
k})$ similarly to $M^{2}$ for $\mathbb{R}^{\ell\times k}$-valued processes.)
\ We adapt It\^{o}'s formula from Li and Peng \cite[Theorem 5.4]{lipeng} or
Soner et al. \cite[Propn. 6.7]{STZ-2010-2} and rewrite it in our context. Let
$0\leq\tau\leq t\leq T$; define $v=\left(  v^{ij}\right)  $ by (\ref{v}).
Then, for any function $f:\mathbb{R}^{d}\rightarrow\mathbb{R}$ with continuous
second order derivatives, we have%
\[
f(X_{t})-f(X_{\tau})=\int\nolimits_{\tau}^{t}(f_{x}(X_{s}))^{\top}\gamma
_{s}dB_{s}+\int\nolimits_{\tau}^{t}(f_{x}(X_{s}))^{\top}\alpha_{s}ds+\tfrac
{1}{2}\int\nolimits_{\tau}^{t}tr[\gamma_{s}^{\top}f_{xx}(X_{s})v_{s}\gamma
_{s}]ds.
\]

Consider the special case $f(x_{1},x_{2})=x_{1}x_{2}$ and
\[
X_{t}^{i}=X_{0}^{i}+\int\nolimits_{0}^{t}\alpha_{s}^{i}ds+\int\nolimits_{0}%
^{t}\gamma_{s}^{i}dB_{s}\text{, }i=1,2\text{,}%
\]
where $\alpha^{i}\in M^{2}$ and $\gamma^{i}\in M^{2}(\mathbb{R}^{d})$,
$i=1,2$. Then%
\[
X_{t}^{1}X_{t}^{2}-X_{\tau}^{1}X_{\tau}^{2}=\int\nolimits_{\tau}^{t}X_{s}%
^{1}dX_{s}^{2}+\int\nolimits_{\tau}^{t}X_{s}^{2}dX_{s}^{1}+\int\nolimits_{\tau
}^{t}\gamma_{s}^{1}v_{s}(\gamma_{s}^{2})^{\top}ds.
\]

\medskip

\noindent\textit{Formal rules:} As in the classical It\^{o} formula, if
$dX_{t}=\alpha_{t}dt+\gamma_{t}dB_{t}$, then we can compute $\ (dX_{t}%
)^{2}=(dX_{t})\cdot(dX_{t})$ \ by the following formal rules:%
\[
dt\cdot dt=dt\cdot dB_{t}=dB_{t}\cdot dt=0\text{, }\;dB_{t}\cdot dB_{t}%
=v_{t}dt.
\]

\medskip

\noindent\textit{Martingale Representation Theorem}: An $\mathcal{F}%
$-progressively measurable $\widehat{L^{2}}(\Omega)$-valued process $X$ is
called a $G$\textit{-martingale} if and only if for any $0\leq\tau<t$,
$X_{\tau}=\hat{E}[X_{t}\mid\mathcal{F}_{\tau}]$. We adapt the martingale
representation theorem from Song \cite{song} and Soner et al. \cite{STZ}. For
any $\xi\in\widehat{L^{2+\varepsilon}}(\Omega)$ and $\varepsilon>0$, if
$X_{t}=\hat{E}[\xi\mid\mathcal{F}_{t}],$ $t\in\lbrack0,T]$, then we have the
following unique decomposition:%
\[
X_{t}=X_{0}+\int\nolimits_{0}^{t}Z_{s}dB_{s}-K_{t}\text{,}%
\]
where $Z\in M^{2}$, $K$ is a continuous nondecreasing process with $K_{0}=0,$
$K_{T}\in\widehat{L^{2}}(\Omega)$\ and where $-K$ is a $G$-martingale.\hfill
\hfill

\section{Appendix: Proofs for Asset Returns\label{app-assetreturns}}

\subsection{Proof of Theorem \ref{thm-hedge}}

\begin{lemma}
Consider the following backward stochastic differential equation (BSDE) driven
by $G$-Brownian motion:%
\begin{align*}
d\tilde{Y}_{t}  & =(r_{t}\tilde{Y}_{t}+\eta_{t}^{\top}\tilde{\phi}_{t}%
-\delta_{t})dt-dK_{t}+\tilde{\phi}_{t}^{\top}dB_{t}\text{,}\;\\
\tilde{Y}_{T}  & =\delta_{T}\text{.}%
\end{align*}
Denote by $I(0,T)$ the space of all continuous nondecreasing processes
$(K_{t})_{0\leq t\leq T}$ with $K_{0}=0$ and $K_{T}\in\widehat{L^{2}}(\Omega
)$. Then there exists a unique triple
\[
(\tilde{Y}_{t},\tilde{\phi}_{t},K_{t})\in M^{2}\times M^{2}\times
I(0,T)\text{,}%
\]
satisfying the BSDE such that $K_{0}=0$ and where $-K_{t}$ is a $G$-martingale.
\end{lemma}%

\textbf{Proof.}
Apply Ito's formula to $\pi_{t}\tilde{Y}_{t}$ to derive%
\[%
\begin{array}
[c]{rl}
& d(\pi_{t}\tilde{Y}_{t})\\
= & \pi_{t}d\tilde{Y}_{t}+\tilde{Y}_{t}d\pi_{t}-\langle\pi_{t}\tilde{\phi}%
_{t}^{\top},\eta_{t}^{\top}v_{t}^{-1}d\langle B\rangle_{t}\rangle\\
= & (\pi_{t}\tilde{\phi}_{t}-\pi_{t}\tilde{Y}_{t}\eta_{t}^{\top}v_{t}%
^{-1})dB_{t}-\pi_{t}\delta_{t}dt-\pi_{t}dK_{t}+[\pi_{t}\tilde{\phi}_{t}^{\top
}\eta_{t}dt-\langle\pi_{t}\tilde{\phi}_{t}^{\top},\eta_{t}^{\top}v_{t}%
^{-1}d\langle B\rangle_{t}\rangle]\\
= & (\pi_{t}\tilde{\phi}_{t}-\pi_{t}\tilde{Y}_{t}\eta_{t}^{\top}v_{t}%
^{-1})dB_{t}-\pi_{t}\delta_{t}dt-\pi_{t}dK_{t}\text{.}%
\end{array}
\]
Integrate on both sides to obtain%
\begin{equation}
\pi_{T}\delta_{T}+%
{\displaystyle\int\nolimits_{\tau}^{T}}
\pi_{t}\delta_{t}dt=\pi_{\tau}\tilde{Y}_{\tau}-%
{\displaystyle\int\nolimits_{\tau}^{T}}
\pi_{t}dK_{t}+%
{\displaystyle\int\nolimits_{\tau}^{T}}
(\pi_{t}\tilde{\phi}_{t}^{\top}-\pi_{t}\tilde{Y}_{t}\eta_{t}^{\top}v_{t}%
^{-1})dB_{t}.\label{solution-BSDE}%
\end{equation}

Let
\[
X_{\tau}=\hat{E}[\pi_{T}\delta_{T}+%
{\displaystyle\int\nolimits_{0}^{T}}
\pi_{t}\delta_{t}dt\mid\mathcal{F}_{\tau}].
\]
Then $(X_{\tau})$ is a $G$-martingale. By the martingale representation
theorem (Appendix \ref{app-GBM}), there exists a unique pair $(Z_{t}%
,\overline{K}_{t})\in M^{2}\times I(0,T)$ such that
\[
X_{\tau}=\hat{E}[\pi_{T}\delta_{T}+%
{\displaystyle\int\nolimits_{0}^{T}}
\pi_{t}\delta_{t}dt]+%
{\displaystyle\int\nolimits_{0}^{\tau}}
Z_{t}dB_{t}-\overline{K}_{t}\text{,}%
\]
and such that $-\overline{K}_{t}$ is a $G$-martingale. This can be rewritten
as
\[%
\begin{array}
[c]{rl}%
X_{\tau} & =X_{T}-%
{\displaystyle\int\nolimits_{\tau}^{T}}
Z_{t}dB_{t}+\overline{K}_{T}-\overline{K}_{t}\\
& =\pi_{T}\delta_{T}+%
{\displaystyle\int\nolimits_{0}^{T}}
\pi_{t}\delta_{t}dt-%
{\displaystyle\int\nolimits_{\tau}^{T}}
Z_{t}dB_{t}+\overline{K}_{T}-\overline{K}_{\tau}.
\end{array}
\]
Thus $(\tilde{Y}_{t},\tilde{\phi}_{t},K_{t})$ is the desired solution where
\[%
\begin{array}
[c]{c}%
\tilde{Y}_{\tau}=\frac{X_{\tau}}{\pi_{\tau}}-%
{\displaystyle\int\nolimits_{0}^{\tau}}
\frac{\pi_{t}}{\pi_{\tau}}\delta_{t}dt\text{,}\\
\tilde{\phi}_{\tau}^{\top}=\frac{Z_{\tau}}{\pi_{\tau}}+\tilde{Y}_{\tau}%
\eta_{\tau}^{\top}v_{t}^{-1}\text{, and }K_{\tau}=%
{\displaystyle\int\nolimits_{0}^{\tau}}
\frac{1}{\pi_{t}}d\overline{K}_{t}.\hfill\text{ }\blacksquare
\end{array}
\hfill
\]
\hfill\hfill\hfill

\medskip\bigskip

Turn to proof of the theorem. We prove only the claim re superhedging. Proof
of the other claim is similar.

\medskip

\noindent\textbf{Step 1:} Prove that for any $y\in\mathcal{U}_{\tau}$,
\[
y\geq\hat{E}[%
{\displaystyle\int\nolimits_{\tau}^{T}}
\frac{\pi_{t}}{\pi_{\tau}}\delta_{t}dt+\frac{\pi_{T}}{\pi_{\tau}}\delta
_{T}\mid\mathcal{F}_{\tau}]
\]

\noindent If $y\in\mathcal{U}_{\tau}$, there exists $\phi$ such that
$Y_{T}^{y,\phi,\tau}\geq\delta_{T}$. Apply (the G-Brownian version of)
It\^{o}'s formula to $\pi_{t}Y_{t}^{y,\phi,\tau}$ to derive%
\[%
\begin{array}
[c]{rl}
& d(\pi_{t}Y_{t}^{y,\phi,\tau})\\
= & \pi_{t}dY_{t}^{y,\phi,\tau}+Y_{t}^{y,\phi,\tau}d\pi_{t}-\langle\pi_{t}%
\phi_{t}^{\top},\eta_{t}^{\top}v_{t}^{-1}d\langle B\rangle_{t}\rangle\\
= & (\pi_{t}\phi_{t}-\pi_{t}Y_{t}^{y,\phi,\tau}\eta_{t}^{\top}v_{t}%
^{-1})dB_{t}-\pi_{t}\delta_{t}dt+[\pi_{t}\phi_{t}^{\top}\eta_{t}dt-\langle
\pi_{t}\phi_{t}^{\top},\eta_{t}^{\top}v_{t}^{-1}d\langle B\rangle_{t}%
\rangle]\text{.}%
\end{array}
\]
Integration on both sides yields%
\[
\pi_{T}Y_{T}^{y,\phi,\tau}+%
{\displaystyle\int\nolimits_{\tau}^{T}}
\pi_{t}\delta_{t}dt=\pi_{\tau}y+%
{\displaystyle\int\nolimits_{\tau}^{T}}
(\pi_{t}\phi_{t}^{\top}-\pi_{t}Y_{t}^{y,\phi,\tau}\eta_{t}^{\top}v_{t}%
^{-1})dB_{t}\text{,}%
\]
and taking conditional expectations yields
\begin{align*}
y  & =\hat{E}[\frac{\pi_{T}}{\pi_{\tau}}Y_{T}^{y,\phi,\tau}+%
{\displaystyle\int\nolimits_{\tau}^{T}}
\frac{\pi_{t}}{\pi_{\tau}}\delta_{t}dt\mid\mathcal{F}_{\tau}]\\
& \geq\hat{E}[\frac{\pi_{T}}{\pi_{\tau}}\delta_{T}+%
{\displaystyle\int\nolimits_{\tau}^{T}}
\frac{\pi_{t}}{\pi_{\tau}}\delta_{t}dt\mid\mathcal{F}_{\tau}]\text{.}%
\end{align*}

\noindent\textbf{Step 2: }There exists $\hat{y}\in\mathcal{U}_{\tau}$ and
$\hat{\phi}$ such that $Y_{T}^{\hat{y},\hat{\phi},\tau}\geq\delta_{T}$ and
\[
\hat{y}=\hat{E}[%
{\displaystyle\int\nolimits_{\tau}^{T}}
\frac{\pi_{t}}{\pi_{\tau}}\delta_{t}dt+\frac{\pi_{T}}{\pi_{\tau}}\delta
_{T}\mid\mathcal{F}_{\tau}]\text{.}%
\]

Apply the preceding lemma. Rewrite equation (\ref{solution-BSDE}) as%
\begin{equation}
\pi_{T}\delta_{T}+%
{\displaystyle\int\nolimits_{\tau}^{T}}
\pi_{t}\delta_{t}dt=\pi_{\tau}\tilde{Y}_{\tau}+%
{\displaystyle\int\nolimits_{\tau}^{T}}
(\pi_{t}\tilde{\phi}_{t}^{\top}-\pi_{t}\tilde{Y}_{t}\eta_{t}^{\top}v_{t}%
^{-1})dB_{t}-%
{\displaystyle\int\nolimits_{\tau}^{T}}
\pi_{t}dK_{t}.\label{super-1}%
\end{equation}
Because $\pi_{t}$ is positive and $-K_{t}$ is a $G$-martingale, $\hat{E}[-%
{\displaystyle\int\nolimits_{\tau}^{T}}
\pi_{t}dK_{t}\mid\mathcal{F}_{\tau}]=0$. Thus,
\[
\hat{E}[\pi_{T}\delta_{T}+%
{\displaystyle\int\nolimits_{\tau}^{T}}
\pi_{t}\delta_{t}dt\mid\mathcal{F}_{\tau}]=\pi_{\tau}\tilde{Y}_{\tau}%
\]
Finally, define $\hat{y}=\tilde{Y}_{\tau}$ and $\hat{\phi}=\tilde{\phi}$. Then
$\hat{y}\in\mathcal{U}_{\tau}$ and
\[
\hat{y}=\hat{E}[%
{\displaystyle\int\nolimits_{\tau}^{T}}
\frac{\pi_{t}}{\pi_{\tau}}\delta_{t}dt+\frac{\pi_{T}}{\pi_{\tau}}\delta
_{T}\mid\mathcal{F}_{\tau}]\text{.}%
\]
This completes the proof of Theorem \ref{thm-hedge}.

\subsection{Proof of Theorem \ref{thm-pi}}

\noindent(i) Apply It\^{o}'s formula for $G$-Brownian motion to
derive\footnote{For any $d$-dimensional (column) vectors $x$ and $y$, we use
$\langle x^{\top},y^{\top}\rangle$ occasionally as alternative notation for
the inner product $x^{\top}y$.}
\begin{equation}%
\begin{array}
[c]{rl}
& d(\pi_{t}Y_{t})\\
= & \pi_{t}dY_{t}+Y_{t}d\pi_{t}-\langle\pi_{t}\phi_{t}^{\top},\eta_{t}^{\top
}v_{t}^{-1}d\langle B\rangle_{t}\rangle\\
= & (\pi_{t}\phi_{t}-\pi_{t}Y_{t}\eta_{t}^{\top}v_{t}^{-1})dB_{t}-\pi
_{t}(c_{t}-e_{t})dt+[\pi_{t}\phi_{t}^{\top}\eta_{t}dt-\langle\pi_{t}\phi
_{t}^{\top},\eta_{t}^{\top}v_{t}^{-1}d\langle B\rangle_{t}\rangle].
\end{array}
\label{g-ito}%
\end{equation}
Note that for any $a=(a_{t})\in M^{2}$,{\Large \ }%
\[
\int\nolimits_{\tau}^{T}a_{t}v_{t}^{-1}d\langle B\rangle_{t}=\int%
\nolimits_{\tau}^{T}a_{t}dt,
\]
and therefore,
\[%
{\displaystyle\int\nolimits_{\tau}^{T}}
\langle\pi_{t}\phi_{t},\eta_{t}^{\top}v_{t}^{-1}d\langle B\rangle_{t}\rangle=%
{\displaystyle\int\nolimits_{\tau}^{T}}
\pi_{t}\phi_{t}^{\top}\eta_{t}dt\text{.}%
\]
Accordingly, integration on both sides of (\ref{g-ito}) yields,
\[
\pi_{T}Y_{T}+%
{\displaystyle\int\nolimits_{\tau}^{T}}
\pi_{t}(c_{t}-e_{t})dt=\pi_{\tau}Y_{\tau}+%
{\displaystyle\int\nolimits_{\tau}^{T}}
(\pi_{t}\phi_{t}^{\top}-\pi_{t}Y_{t}\eta_{t}^{\top}v_{t}^{-1})dB_{t}.
\]
Take conditional expectations to obtain
\[
\hat{E}[\pi_{T}Y_{T}+%
{\displaystyle\int\nolimits_{\tau}^{T}}
\pi_{t}(c_{t}-e_{t})dt\mid\mathcal{F}_{\tau}]=\pi_{\tau}Y_{\tau}+\hat{E}[%
{\displaystyle\int\nolimits_{\tau}^{T}}
(\pi_{t}\phi_{t}^{\top}-\pi_{t}Y_{t}\eta_{t}^{\top}v_{t}^{-1})dB_{t}%
\mid\mathcal{F}_{\tau}]\text{.}%
\]
Because $B$ being $G$-Brownian motion implies that $B$ is a martingale under
every prior in $\mathcal{P}$, we have
\[
0=\hat{E}[%
{\displaystyle\int\nolimits_{\tau}^{T}}
(\pi_{t}\phi_{t}^{\top}-\pi_{t}Y_{t}\eta_{t}^{\top}v_{t}^{-1})dB_{t}%
\mid\mathcal{F}_{\tau}]=\hat{E}[-%
{\displaystyle\int\nolimits_{\tau}^{T}}
(\pi_{t}\phi_{t}^{\top}-\pi_{t}Y_{t}\eta_{t}^{\top}v_{t}^{-1})dB_{t}%
\mid\mathcal{F}_{\tau}]\text{,}%
\]
which gives the desired result.

\bigskip

\noindent(ii) We need to find a process $\phi$\ such that, for the given $c
$,\ the solution $(Y_{t})$ to%
\begin{align*}
dY_{t}  & =(r_{t}Y_{t}+\eta_{t}^{\top}\phi_{t}-(c_{t}-e_{t}))dt+\phi_{t}%
^{\top}dB_{t}\text{, ~}t\in\lbrack\tau,T]\\
Y_{T}  & =c_{T}-e_{T}%
\end{align*}
has time $\tau$\ wealth equal to the given value $Y_{\tau}$.

For $\tau\leq s\leq T$, define
\[
X_{s}\equiv\hat{E}[%
{\displaystyle\int\nolimits_{\tau}^{T}}
\frac{\pi_{t}}{\pi_{\tau}}(c_{t}-e_{t})dt+\frac{\pi_{T}}{\pi_{\tau}}%
(c_{T}-e_{T})\mid\mathcal{F}_{s}].
\]
Then $X_{s}=-\hat{E}[-%
{\displaystyle\int\nolimits_{\tau}^{T}}
\frac{\pi_{t}}{\pi_{\tau}}(c_{t}-e_{t})dt-\frac{\pi_{T}}{\pi_{\tau}}%
(c_{T}-e_{T})\mid\mathcal{F}_{s}]$ and $(X_{s})_{\tau\leq s\leq T}$ is a
symmetric G-martingale. By Soner et. al. \cite{STZ} and Song \cite{song}, it
admits the unique representation%

\[
X_{s}=X_{\tau}+\int_{\tau}^{s}Z_{t}^{\top}dB_{t}\text{, }%
\]
where $Z\in$ $M^{2}$. Note that%
\[
X_{\tau}=\hat{E}[%
{\displaystyle\int\nolimits_{\tau}^{T}}
\frac{\pi_{t}}{\pi_{\tau}}(c_{t}-e_{t})dt+\frac{\pi_{T}}{\pi_{\tau}}%
(c_{T}-e_{T})\mid\mathcal{F}_{\tau}]=Y_{\tau}.
\]

Set
\[
\bar{Y}_{s}\equiv X_{s}-%
{\displaystyle\int\nolimits_{\tau}^{s}}
\frac{\pi_{t}}{\pi_{\tau}}(c_{t}-e_{t})dt,\text{~}s\in\lbrack\tau,T].
\]
Then $(\bar{Y}_{s})$ satisfies%
\[
d\bar{Y}_{s}=-\frac{\pi_{s}}{\pi_{\tau}}(c_{s}-e_{s})ds+Z_{s}^{\top}%
dB_{s}\text{, \ }\bar{Y}_{\tau}=Y_{\tau}.
\]
Define%

\[
Y_{s}\equiv\bar{Y}_{s}(\frac{\pi_{s}}{\pi_{\tau}})^{-1}.
\]
Note that $(\pi_{s})$ satisfies%
\[
d\pi_{s}/\pi_{s}=-r_{s}ds-\eta_{s}^{\top}v_{s}^{-1}dB_{s}\text{,~}s\in
\lbrack\tau,T].
\]
Apply Ito's formula for $G$-Brownian motion to derive%
\[
dY_{s}=[r_{s}Y_{s}+\eta_{s}^{\top}(Y_{s}(v_{s}^{-1})^{\top}\eta_{s}+\frac
{\pi_{\tau}}{\pi_{s}}Z_{s})-(c_{s}-e_{s})]ds+(Y_{s}\eta_{s}^{\top}v_{s}%
^{-1}+\frac{\pi_{\tau}}{\pi_{s}}Z_{s}^{\top})dB_{s}\text{.}%
\]
Finally, set%
\[
\phi_{s}^{\top}\equiv Y_{s}\eta_{s}^{\top}v_{s}^{-1}+\frac{\pi_{\tau}}{\pi
_{s}}Z_{s}^{\top}.
\]
Then%
\[
dY_{s}=(r_{s}Y_{s}+\eta_{s}^{\top}\phi_{t}-(c_{s}-e_{s}))ds+\phi_{s}^{\top
}dB_{s}\text{, ~}s\in\lbrack\tau,T].
\]
This completes the proof.\hfill$\blacksquare$

\subsection{Proof of Theorem \ref{thm-eqI}}

The proof follows from Theorem \ref{thm-pi} and the following lemma.

\begin{lemma}
For every $c$, we have: For each $\tau$, $\mathcal{P}$-almost surely,
\begin{align}
\widehat{E}\left[  \int_{\tau}^{T}\pi_{t}^{e}\left(  c_{t}-e_{t}\right)
dt+\pi_{T}^{e}\left(  c_{T}-e_{T}\right)  \mid\mathcal{F}_{\tau}\right]   &
\leq0\Longrightarrow\label{budget2}\\
V_{\tau}\left(  c\right)   & \leq V_{\tau}\left(  e\right)  \text{.}\nonumber
\end{align}

\end{lemma}%

\textbf{Proof.}
Define $\delta_{t}$ implicitly by%
\[
f(c_{t},V_{t}(c))=f_{c}(e_{t},V_{t}\left(  e\right)  )(c_{t}-e_{t}%
)+f_{v}(e_{t},V_{t}\left(  e\right)  )(V_{t}(c)-V_{t}(e))-\delta_{t}%
+f(e_{t},V_{t}\left(  e\right)  )\text{,}%
\]
for $0\leq t<T$, and%
\[
u(c_{T})=u_{c}(e_{T})(c_{T}-e_{T})-\delta_{T}+u(e_{T})\text{.}%
\]
Because $f$ and $u$ are concave, we have $\ \delta_{t}\geq0$ \ on $\left[
0,T\right]  $.

\noindent Define, for $0\leq t<T$,
\[%
\begin{array}
[c]{l}%
\beta_{t}=f_{v}(e_{t},V_{t}\left(  e\right)  )\\
\gamma_{t}=f_{c}(e_{t},V_{t}\left(  e\right)  )(c_{t}-e_{t})+f(e_{t}%
,V_{t}\left(  e\right)  )-\beta_{t}V_{t}(e)-\delta_{t}\\
\gamma_{T}=-u_{c}(e_{T})e_{T}+u(e_{T})-\delta_{T}\\
\zeta_{t}=f(e_{t},V_{t}\left(  e\right)  )-\beta_{t}V_{t}(e)\text{.}%
\end{array}
\]
Then%

\[
V_{t}(c)=-\hat{E}[-(u_{c}(e_{T})c_{T}+\gamma_{T})-\int_{t}^{T}(\beta_{s}%
V_{s}(c)+\gamma_{s})ds\mid\mathcal{F}_{t}].
\]
Because this is a linear backward stochastic differential equation, its
solution has the form (by Hu and Ji \cite{HuShaolin})
\[
V_{t}(c)=-\hat{E}[-(u_{c}(e_{T})c_{T}+\gamma_{T})\exp\{\int_{t}^{T}\beta
_{s}ds\}-\int_{t}^{T}\gamma_{s}\exp\{\int_{t}^{s}\beta_{s^{\prime}}ds^{\prime
}\}ds\mid\mathcal{F}_{t}]\text{.}%
\]
Similarly for $e$, we have%
\[
V_{t}(e)=-\hat{E}[-u(e_{T})-\int_{t}^{T}(\beta_{s}V_{s}(e)+\zeta_{s}%
)ds\mid\mathcal{F}_{t}]\text{,}%
\]
and (by Hu and Ji \cite{HuShaolin}),%
\[
V_{t}(e)=-\hat{E}[-u(e_{T})\exp\{\int_{t}^{T}\beta_{s}ds\}-\int_{t}^{T}%
\zeta_{s}\exp\{\int_{t}^{s}\beta_{s^{\prime}}ds^{\prime}\}ds\mid
\mathcal{F}_{t}].
\]

Apply the subadditivity of $\hat{E}\left[  \cdot\mid\mathcal{F}_{\tau}\right]
$ and the nonnegativity of $\delta_{t}$ to obtain%
\[%
\begin{array}
[c]{l}%
\exp\{\int_{0}^{\tau}\beta_{s}ds\}\left(  V_{\tau}(c)-V_{\tau}(e)\right)  =\\%
\begin{array}
[c]{l}%
-\hat{E}[-(u_{c}(e_{T})c_{T}+\gamma_{T})\exp\{\int_{0}^{T}\beta_{s}ds\}-%
{\displaystyle\int\nolimits_{\tau}^{T}}
\gamma_{t}\exp\{\int_{0}^{t}\beta_{s}ds\}dt\mid\mathcal{F}_{\tau}]\\
-\{-\hat{E}[-u(e_{T})\exp\{\int_{0}^{T}\beta_{s}ds\}-%
{\displaystyle\int\nolimits_{\tau}^{T}}
\zeta_{t}\exp\{\int_{0}^{t}\beta_{s}ds\}dt\mid\mathcal{F}_{\tau}]\}=
\end{array}
\end{array}
\]%
\[%
\begin{array}
[c]{l}%
\begin{array}
[c]{l}%
\hat{E}[-u(e_{T})\exp\{\int_{0}^{T}\beta_{s}ds\}-%
{\displaystyle\int\nolimits_{\tau}^{T}}
\zeta_{t}\exp\{\int_{0}^{t}\beta_{s}ds\}dt\mid\mathcal{F}_{\tau}]\\
-\hat{E}[-(u_{c}(e_{T})c_{T}+\gamma_{T})\exp\{\int_{0}^{T}\beta_{s}ds\}-%
{\displaystyle\int\nolimits_{\tau}^{T}}
\gamma_{t}\exp\{\int_{0}^{t}\beta_{s}ds\}dt\mid\mathcal{F}_{\tau}]\leq
\end{array}
\\%
\begin{array}
[c]{l}%
\hat{E}[-u(e_{T})\exp\{\int_{0}^{T}\beta_{s}ds\}-%
{\displaystyle\int\nolimits_{\tau}^{T}}
\zeta_{t}\exp\{\int_{0}^{t}\beta_{s}ds\}dt\\
-(-(u_{c}(e_{T})c_{T}+\gamma_{T})\exp\{\int_{0}^{T}\beta_{s}ds\}-%
{\displaystyle\int\nolimits_{\tau}^{T}}
\gamma_{t}\exp\{\int_{0}^{t}\beta_{s}ds\}dt\mid\mathcal{F}_{\tau}]=
\end{array}
\\%
\begin{array}
[c]{l}%
\hat{E}[(u_{c}(e_{T})c_{T}+\gamma_{T})\exp\{\int_{0}^{T}\beta_{s}ds\}+%
{\displaystyle\int\nolimits_{\tau}^{T}}
\gamma_{t}\exp\{\int_{0}^{t}\beta_{s}ds\}dt\\
-u(e_{T})\exp\{\int_{0}^{T}\beta_{s}ds\}-%
{\displaystyle\int\nolimits_{\tau}^{T}}
\zeta_{t}\exp\{\int_{0}^{t}\beta_{s}ds\}dt\mid\mathcal{F}_{\tau}]=
\end{array}
\end{array}
\]%
\[%
\begin{array}
[c]{l}%
\hat{E}[\exp\{\int_{0}^{T}\beta_{s}ds\}u_{c}(e_{T})(c_{T}-e_{T})+%
{\displaystyle\int\nolimits_{\tau}^{T}}
\exp\{\int_{0}^{t}\beta_{s}ds\}f_{c}(e_{t},V_{t}\left(  e\right)
)(c_{t}-e_{t})dt\\
-\exp\{\int_{0}^{T}\beta_{s}ds\}\delta_{T}-%
{\displaystyle\int\nolimits_{\tau}^{T}}
\exp\{\int_{0}^{t}\beta_{s}ds\}\delta_{t}dt\mid\mathcal{F}_{\tau}]\leq\\%
\begin{array}
[c]{l}%
\hat{E}[\exp\{\int_{0}^{T}\beta_{s}ds\}u_{c}(e_{T})(c_{T}-e_{T})+%
{\displaystyle\int\nolimits_{\tau}^{T}}
\exp\{\int_{0}^{t}\beta_{s}ds\}f_{c}(e_{t},V_{t}\left(  e\right)
)(c_{t}-e_{t})dt\mid\mathcal{F}_{\tau}]\\
=\widehat{E}\left[  \pi_{T}^{e}\left(  c_{T}-e_{T}\right)  +\int_{\tau}^{T}%
\pi_{t}^{e}\left(  c_{t}-e_{t}\right)  dt\mid\mathcal{F}_{\tau}\right]
\text{.}%
\end{array}
\end{array}
\]
This completes the proof. \hfill$\blacksquare$

\subsection{C-CAPM for General Aggregators\label{app-KP}}

We derive (\ref{returnsKP}) and the corresponding form of the C-CAPM for
general aggregators, thus justifying claims made following Theorem
\ref{thm-eqI}. Utility is defined by (\ref{Vt-general}).

\begin{lemma}
For given $c\in D$, \ there is a unique solution $V_{t}$ to
\[
V_{t}\left(  c\right)  =-\hat{E}[-\int_{t}^{T}f(c_{s},V_{s}\left(  c\right)
)ds-u\left(  c_{T}\right)  \mid\mathcal{F}_{t}],~0\leq t\leq T.
\]
Further, there exist unique $Z\in M^{2}$ and $K$ (a continuous nondecreasing
process with $K_{0}=0$) such that%
\[
V_{t}=u\left(  c_{T}\right)  +\int_{t}^{T}f(c_{s},V_{s}\left(  c\right)
)ds+\int_{t}^{T}Z_{s}dB_{s}-K_{T}+K_{t}.
\]

\end{lemma}%

\textbf{Proof.}
Define%
\[
U_{t}=-\hat{E}[-\int_{0}^{T}f(c_{s},V_{s}\left(  c\right)  )ds-u\left(
c_{T}\right)  \mid\mathcal{F}_{t}]\text{, ~}0\leq t\leq T\text{.}%
\]
Note that
\begin{align*}
U_{0}  & =-\hat{E}[-\int_{0}^{T}f(c_{s},V_{s}\left(  c\right)  )ds-u\left(
c_{T}\right)  ]\text{,}\\
U_{T}  & =\int_{0}^{T}f(c_{s},V_{s}\left(  c\right)  )ds-u\left(
c_{T}\right)  \text{.}%
\end{align*}
Because $-U_{t}$ is a G-martingale, it has the following unique
representation:%
\[
-U_{t}=-U_{0}+\int_{0}^{t}Z_{s}dB_{s}-K_{t}\text{.}%
\]
Then
\begin{align*}
V_{t}  & =-\hat{E}[-\int_{t}^{T}f(c_{s},V_{s}\left(  c\right)  )ds-u\left(
c_{T}\right)  \mid\mathcal{F}_{t}]\\
& =U_{t}-\int_{0}^{t}f(c_{s},V_{s}\left(  c\right)  )ds\\
& =U_{0}-\int_{0}^{t}Z_{s}dB_{s}+K_{t}-\int_{0}^{t}f(c_{s},V_{s}\left(
c\right)  )ds.
\end{align*}

Note that%
\begin{align*}
V_{T}  & =u\left(  c_{T}\right) \\
& =U_{0}-\int_{0}^{T}Z_{s}dB_{s}+K_{T}-\int_{0}^{T}f(c_{s},V_{s}\left(
c\right)  )ds\Longrightarrow\\
V_{t}-V_{T}  & =V_{t}-u\left(  c_{T}\right) \\
& =\int_{t}^{T}f(c_{s},V_{s}\left(  c\right)  )ds+\int_{t}^{T}Z_{s}%
dB_{s}-K_{T}+K_{t}\Longrightarrow\\
V_{t}  & =u\left(  c_{T}\right)  +\int_{t}^{T}f(c_{s},V_{s}\left(  c\right)
)ds+\int_{t}^{T}Z_{s}dB_{s}-K_{T}+K_{t}\text{.}%
\end{align*}
Uniqueness of $\left(  V_{t}\right)  $ follows by standard contraction mapping
arguments (see our companion paper). \hfill$\blacksquare$

\bigskip

The preceding representation of utility, combined with Ito's Lemma for
$G$-Brownian motion, yields
\begin{equation}
b_{t}-r_{t}1=-\frac{f_{cc}\left(  e_{t},V_{t}\right)  e_{t}}{f_{c}\left(
e_{t},V_{t}\right)  }s_{t}v_{t}s_{t}^{e}+\frac{f_{cV}\left(  e_{t}%
,V_{t}\right)  }{f_{c}\left(  e_{t},V_{t}\right)  }s_{t}v_{t}Z_{t}^{\top
}\text{.}\label{g-capm}%
\end{equation}
For the Kreps-Porteus aggregator (\ref{KPagg}), this becomes
\begin{equation}
b_{t}-r_{t}1=(1-\rho)s_{t}v_{t}s_{t}^{e}+\frac{\alpha-\rho}{\alpha}s_{t}%
v_{t}\frac{Z_{t}^{\top}}{V_{t}}\text{.}\label{zz}%
\end{equation}
Then (\ref{returnsKP}) follows from the following relation (which can be
proven as in Chen and Epstein \cite{CE}{\LARGE \ }by exploiting the
homogeneity of degree $\alpha$ of utility):
\[
Z_{t}/(\alpha V_{t})=\rho^{-1}[s^{M}+(\rho-1)s_{t}^{e}]\text{.}%
\]

\subsection{Proof of Theorem \ref{thm-eqII}}

The strategy is to argue that for any $c\in \Upsilon_{\tau}\left(  0\right)
$,
\begin{align*}
V_{\tau}\left(  c\right)  -V_{\tau}\left(  e\right)   & =V_{\tau}\left(
c\right)  -E^{(P^{\ast})_{\tau}^{\omega}}[\int_{\tau}^{T}f(e_{s},V_{s}\left(
e\right)  )ds+u\left(  e_{T}\right)  ]\\
^{\dagger}  & \leq E^{(P^{\ast})_{\tau}^{\omega}}[\int_{\tau}^{T}f(c_{s}%
,V_{s}\left(  c\right)  )ds+u\left(  c_{T}\right)  ]\\
& -E^{(P^{\ast})_{\tau}^{\omega}}[\int_{\tau}^{T}f(e_{s},V_{s}\left(
e\right)  )ds+u\left(  e_{T}\right)  ]\\
& \leq E^{(P^{\ast})_{\tau}^{\omega}}[\int_{\tau}^{T}\pi_{t}^{e}(c_{t}%
-e_{t})dt+\pi_{T}^{e}\left(  c_{T}-e_{T}\right)  ]
\end{align*}%
\begin{align*}
\text{by (\ref{pipie2})}  & =f_{c}\left(  e_{0},V_{0}\left(  e\right)
\right)  E^{(P^{\ast})_{\tau}^{\omega}}[\int_{\tau}^{T}\pi_{t}(c_{t}%
-e_{t})dt+\pi_{T}\left(  c_{T}-e_{T}\right)  ]\\
^{\dagger\dagger}  & \leq f_{c}\left(  e_{0},V_{0}\left(  e\right)  \right)
\widehat{E}[\int_{\tau}^{T}\pi_{t}(c_{t}-e_{t})dt+\pi_{T}\left(  c_{T}%
-e_{T}\right)  \mid\mathcal{F}_{\tau}]\leq0\text{.}%
\end{align*}
The inequalities marked $\dagger$ and $\dagger\dagger$ are justified in the
next lemma.

\begin{lemma}
\label{lemma-eq}For every $\tau$, $\mathcal{P}$-almost surely,
\begin{equation}
V_{\tau}\left(  c\right)  \leq E^{\left(  P^{\ast}\right)  _{\tau}^{\omega}%
}[\int_{\tau}^{T}f(e_{s},V_{s}\left(  e\right)  )ds+u\left(  e_{T}\right)
]\text{,}\label{Vineq}%
\end{equation}
and%
\begin{align*}
& E^{\left(  P^{\ast}\right)  _{\tau}^{\omega}}[\int_{\tau}^{T}\pi_{t}%
(c_{t}-e_{t})dt+\pi_{T}\left(  c_{T}-e_{T}\right)  ]\\
& \leq\widehat{E}[\int_{\tau}^{T}\pi_{t}(c_{t}-e_{t})ds+\pi_{T}\left(
c_{T}-e_{T}\right)  \mid\mathcal{F}_{\tau}]\text{.}%
\end{align*}

\end{lemma}%

\textbf{Proof.}
We prove the first inequality. The second is proven similarly.

We claim that for any $P\in\mathcal{P}$ and $\tau$, there exists $\overline
{P}\in\mathcal{P}$ such that%
\begin{equation}
\overline{P}=P\text{ on }\mathcal{F}_{\tau}\text{ and }\overline{P}%
_{t}^{\omega}=\left(  P^{\ast}\right)  _{t}^{\omega}\text{ for all }\left(
t,\omega\right)  \in\lbrack\tau,T]\times\Omega\text{.}\label{pasting}%
\end{equation}
This follows from the construction of priors in $\mathcal{P}$ via the SDE
(\ref{Xtheta}). Let $P^{\ast}$ and $P$ be induced by $\theta^{\ast}$ and
$\theta$ respectively and define $\overline{\theta}\in\Theta$ by%
\[
\overline{\theta}_{t}=\left\{
\begin{array}
[c]{cc}%
\theta_{t} & 0\leq t\leq\tau\\
\theta_{t}^{\ast} & \tau<t\leq T\text{.}%
\end{array}
\right.  \;
\]
Then $\overline{P}=P^{\overline{\theta}}$ satisfies (\ref{pasting}). It
follows from the detailed construction of conditional expectation
$\widehat{E}\left[  \cdot\mid\mathcal{F}_{\tau}\right]  $, that $P$-$a.e.$%
\begin{align*}
V_{\tau}\left(  c\right)   & \leq E^{\overline{P}}[\int_{\tau}^{T}%
f(c_{s},V_{s}\left(  c\right)  )ds+u\left(  c_{T}\right)  \mid\mathcal{F}%
_{\tau}]\text{\ \ \ }\\
& =E^{\overline{P}_{\tau}^{\omega}}[\int_{\tau}^{T}f(c_{s},V_{s}\left(
c\right)  )ds+u\left(  c_{T}\right)  ]\\
& =E^{\left(  P^{\ast}\right)  _{\tau}^{\omega}}[\int_{\tau}^{T}f(c_{s}%
,V_{s}\left(  c\right)  )ds+u\left(  c_{T}\right)  ]
\end{align*}
The first equality follows from $\overline{P}=P$ on $\mathcal{F}_{\tau}$ and
properties of regular conditionals (see Yong and Zhou \cite[Propns. 1.9,
1.10]{YZ}). Moreover, the preceding is true for any $P\in\mathcal{P}$.
\ \hfill$\blacksquare$

\end{document}